\documentclass[12pt,a4paper]{article}

\usepackage{rotating}
\usepackage{amsmath}
\usepackage{mathtools}
\usepackage{amsthm}
\usepackage{amsfonts}
\usepackage[dvips]{epsfig}
\usepackage{graphicx}
\usepackage{amssymb}
\usepackage{cancel}
\usepackage{lscape}
\usepackage{cancel}
\usepackage{slashed}
\usepackage{lscape}
\usepackage{color}
\usepackage{colortbl}
\usepackage{longtable}
\usepackage{subfigure}
\usepackage{cite}
\usepackage{bm}
\usepackage{multirow}
\usepackage{url}

\definecolor{Orange}{cmyk}{0,0.61,0.87,0}
\definecolor{JungleGreen}{cmyk}{0.99,0,0.52,0}
\definecolor{OliveGreen}{cmyk}{0.64,0,0.95,0.40}
\definecolor{Brown}{cmyk}{0,0.70,1,0.40}
\definecolor{RoyalBlue}{cmyk}{0.71,0.53,0,0.12}
\definecolor{Gray}{cmyk}{0,0,0,0.40}
\definecolor{LightPink}{cmyk}{0.0,0.25,0,0}
\definecolor{LLightPink}{cmyk}{0.0,0.10,0,0}
\definecolor{LightBlue}{cmyk}{0.25,0,0,0}
\definecolor{LightGray}{cmyk}{0,0,0,0.2}

\newcommand{\Slash}[1]{{\ooalign{\hfil/\hfil\crcr$#1$}}}

\newcommand{\model}[3]{${\tt #1}_{\tt\bf #2}^{\tt #3}$}
\newcommand{\DM}[2]{${\tt #1}_{\tt #2}$}
\newcommand{\bhline}[1]{\noalign{\hrule height #1}}

\begin{document}

\begin{titlepage}

\begin{flushright}

FTPI--MINN--15/40 \\
UMN--TH--3501/15 \\
IPMU15--0142 \\

\end{flushright}

\vskip 1cm
\begin{center}

{\Large
{\bf 
Weakly-Interacting Massive Particles in \\[3pt]
Non-supersymmetric SO(10) Grand Unified Models
}
}

\vskip 1cm

Natsumi Nagata$^{a,b}$, 
Keith A. Olive$^a$, 
and
Jiaming Zheng$^a$

\vskip 0.5cm

{\it $^{a}$William I. Fine Theoretical Physics Institute, School of
 Physics and Astronomy, University of Minnesota, Minneapolis, MN 55455,
 USA}\\[2pt]
{\it $^{b}$Kavli IPMU (WPI), UTIAS, University of Tokyo,
 Kashiwa, Chiba 277-8583, Japan}
\date{\today}

\vskip 1.5cm

\begin{abstract}

 Non-supersymmetric SO(10) grand unified theories provide a framework in
 which the stability of dark matter is explained while gauge coupling
 unification is realized. In this work, we systematically study this
 possibility by classifying weakly interacting dark matter candidates in terms of
 their quantum numbers of $\text{SU}(2)_L \otimes \text{U}(1)_Y$, $B-L$, and
 $\text{SU}(2)_R$. We consider both scalar and fermion candidates. We show that the
 requirement of a sufficiently high unification scale to ensure a proton
 lifetime compatible with experimental constraints plays a strong role
 in selecting viable candidates. Among the scalar candidates originating
 from either a {\bf 16} or {\bf 144} of SO(10), only SU(2)$_L$ singlets
 with zero hypercharge or doublets with $Y=1/2$ satisfy all constraints
 for $\text{SU}(4)_C \otimes \text{SU}(2)_L \otimes \text{SU}(2)_R$ and
 $\text{SU}(3)_C \otimes \text{SU}(2)_L \otimes \text{SU}(2)_R \otimes
 \text{U}(1)_{B-L}$ intermediate scale gauge groups. Among fermion
 triplets with zero hypercharge, only a triplet in the {\bf 45} with
 intermediate group $\text{SU}(4)_C \otimes \text{SU}(2)_L \otimes
 \text{SU}(2)_R$ leads to solutions with $M_{\rm GUT} > M_{\rm int}$ and
 a long proton lifetime. We find three models with weak doublets and
 $Y=1/2$ as dark matter candidates for the $\text{SU}(4)_C \otimes
 \text{SU}(2)_L \otimes \text{SU}(2)_R$ and $\text{SU}(4)_C \otimes
 \text{SU}(2)_L \otimes \text{U}(1)_R$ intermediate scale gauge
 groups assuming a minimal Higgs content. We also
 discuss how these models may be tested at accelerators and in dark
 matter detection experiments.

\end{abstract}

\end{center}
\end{titlepage}

\section{Introduction}
\label{sec:introduction}

Various cosmological observations have now established that more than
80\% of the energy density of matter in the Universe is composed of
non-baryonic dark matter (DM) \cite{Ade:2015xua}. The Standard Model
(SM) of particle physics, however, cannot explain this observation, and
therefore there should be new physics beyond the SM which contains
a DM candidate. One of the most promising class of candidates for DM is the
so-called weakly-interacting massive particle (WIMP). These are
electrically neutral and colorless particles which have masses of ${\cal
O}(10^{(2-3)})$~GeV and couple to SM particles via weak-scale
interactions. Their thermal relic abundance can explain the current
energy density of DM. Such particles are predicted in many new-physics models;
for example, the lightest neutralino in the supersymmetric (SUSY) SM is
a well-known candidate for WIMP DM \cite{EHNOS}.

For a WIMP to be DM, it should be stable or have a sufficiently long
lifetime compared to the age of the Universe. To assure that, it is
usually assumed that there is a symmetry which stabilizes the DM 
particle. For instance, in the minimal SUSY SM (MSSM), $R$-parity makes the
lightest SUSY particle stable and thus a candidate for DM in the Universe
\cite{EHNOS}. Similarly,  Kaluza-Klein parity in universal extra
dimensional models \cite{Appelquist:2000nn} and $T$-parity in the
Littlest Higgs model \cite{ArkaniHamed:2001nc} yield stable particles,
which can also be promising DM candidates. The ultraviolet (UV) origin
of such a symmetry is, however, often obscure; thus it would be quite
interesting if a theory which offers a DM candidate and simultaneously
explains its stability can be realized as a UV completion rather than
introducing the additional symmetry by hand.

In fact, grand unified theories (GUTs) can provide such a
framework. Suppose that the rank of a GUT gauge group is
larger than four. In this case, the GUT symmetry contains extra
symmetries beyond the SM gauge symmetry. These extra symmetries
should be spontaneously broken at a high-energy scale by a vacuum expectation
value (VEV) of a Higgs field. Then, if we choose the
proper representation for the Higgs field, there remains discrete
symmetries, which can be used for DM stabilization
\cite{Kibble:1982ae, Krauss:1988zc, Ibanez:1991hv,
Martin:1992mq,Kadastik:2009dj,Frigerio:2009wf}.  
The discrete charge of each representation is uniquely determined, and
thus we can systematically identify possible DM candidates for each
symmetry.

In this work, we discuss the concrete realization of this scenario in
non-SUSY SO(10) GUT models. It is widely known that SO(10) GUTs
\cite{Georgi:1974my,so10-2,GN2} have a lot of attractive
features. Firstly, all of the SM quarks and leptons, as well as
right-handed neutrinos, can be embedded into ${\bf 16}$ representations
of SO(10). Secondly, the anomaly cancellation in the SM is naturally
explained since SO(10) is free from anomalies. Thirdly, one obtains
improved gauge coupling unification \cite{GN2,masiero, ssw,
delAguila:1980at,Mohapatra:1982aq,Rajpoot:1980xy,Fukugita:1993fr,DiLuzio:2011my} and
improved fermion mass ratios \cite{GN2,lsw} if partial unification is
achieved at an intermediate mass scale. In addition,
since right-handed neutrinos have masses of the order of the
intermediate scale, small neutrino masses can be explained via the
seesaw mechanism \cite{Minkowski:1977sc} if the intermediate scale is
sufficiently high. SO(10) includes an additional U(1) symmetry, which is
assumed to be broken at the intermediate scale. If the 
Higgs field that breaks this additional U(1) symmetry belongs to a ${\bf
126}$ dimensional 
representation, then a discrete $\mathbb{Z}_2$
symmetry is preserved at low energies. One also finds that as
long as we focus on relatively small representations ($ \le \bf{210}$),
the ${\bf 126}$ Higgs field  leaving a $\mathbb{Z}_2$
symmetry is the only possibility for a
discrete symmetry \cite{DeMontigny:1993gy, Mambrini:2015vna}. We
focus on this case in the following discussion. 

DM candidates appearing in such models can be classified into two types; one
class of DM particles have effectively weak-scale interactions with the SM particles so
that they are thermalized in the early universe, while the other class contains SM singlets which are
never brought into thermal equilibrium. In the latter case, DM particles are produced out of 
equilibrium via the thermal scattering involving 
heavy (intermediate scale) particle exchange processes. This type of DM is
called Non-Equilibrium Thermal DM (NETDM) \cite{Mambrini:2013iaa},
whose realization in SO(10) GUTs was thoroughly discussed in
Ref.~\cite{Mambrini:2015vna}. NETDM is necessarily fermionic as scalar DM
would naturally couple to the SM Higgs bosons. 
Depending on the choice of the intermediate-scale gauge group, candidates for 
NETDM may originate from several
different SO(10) representations such as ${\bf 45, 54, 126}$ or ${\bf 210}$. 
Although the NETDM candidate itself does not affect the running of the gauge couplings
from the weak scale to the intermediate scale, part of the original SO(10) multiplet has a mass
at the intermediate scale and does affect the running up to the GUT scale.
Demanding gauge coupling unification with a GUT scale above $10^{15}$ GeV leaves us with
a limited set of potential NETDM candidates. When we further demand the splitting of the
GUT scale or intermediate scale multiplets so that only a singlet survives at low energies
only two candidates were left: in $\text{SU}(4)_C \otimes \text{SU}(2)_L
\otimes\text{SU}(2)_R$ notations
there were  the Dirac ($\bf{1,1,3}$) originating in a ${\bf 45}$ or a
Weyl ($\bf{15,1,1}$) also originating from a ${\bf 45}$ or a ${\bf 210}$
in SO(10).

In this paper, we study the remaining
possibility, namely, that of WIMP DM candidates in SO(10) GUT models. We
systematically classify WIMP DM candidates in terms of their quantum
numbers and embed them in SO(10) representations. As noted above and discussed in
Ref.~\cite{Mambrini:2015vna}, the presence of DM multiplets
significantly affects the running of the gauge coupling constants. In this case, since the DM candidates
are no longer SM singlets, the running of the gauge couplings is also affected between 
the weak and intermediate scales and 
may spoil gauge coupling unification realized in non-SUSY SO(10)
GUTs. We list WIMP DM models in which gauge coupling unification is
achieved with appropriate GUT and intermediate scales. Then, we study
the phenomenology of these models, such as the relic abundance of DM, the DM
direct detection rate, the
proton decay lifetime, and neutrino masses. It is found that the
condition of gauge coupling unification, as well as the proton decay
bounds, severely restricts the WIMP DM models in SO(10) GUTs. Still,
we obtain some promising candidates, which can be probed in future DM
searches and proton decay experiments. 

This paper is organized as follows.
In the next section, we show the
model setup for the SO(10) WIMP DM scenario. The realization of a
$\mathbb{Z}_2$ symmetry and the classification of WIMP DM candidates are
discussed there. 
Then, we analyze the scalar and fermionic DM models in
Sec.~\ref{sec:scalarDM} and Sec.~\ref{sec:fermionDM},
respectively. Section~\ref{sec:conclusion} is devoted to conclusion and
discussion.

\section{Model}

We begin with an overview of the basic SO(10) model needed to accommodate  a DM candidate.
As mentioned above, in this work, we consider SO(10) GUT models
and restrict ourselves to a two step simultaneous symmetry breaking
chain,\footnote{For recent work on this kind of SO(10) scenario, see
Ref.~\cite{Babu:2015bna}. } in which the SO(10) gauge group is broken to an intermediate
gauge group $G_{\text{int}}$ at the GUT scale $M_{\text{GUT}}$, and
subsequently broken to the SM gauge group
$G_\text{SM}\equiv\text{SU(3)}_C\otimes\text{SU(2)}_L\otimes\text{U(1)}_Y$
and a $\mathbb{Z}_2$ symmetry at the intermediate scale
$M_{\text{int}}$:  
\begin{equation}
 \text{SO}(10)\longrightarrow G_{\text{int}}\longrightarrow
 G_{\text{SM}}\otimes \mathbb{Z}_2 ~,
\label{eq:decaypattern}
\end{equation}
The Higgs multiplets that break SO(10) and $G_{\text{int}}$ are labeled
by $R_1$ and $R_2$, respectively. As discussed in the introduction, this
$\mathbb{Z}_2$ symmetry is a remnant of an extra U(1) symmetry in
SO(10) \cite{Kibble:1982ae, Krauss:1988zc, Ibanez:1991hv,
Martin:1992mq} and is used to stabilize DM candidates
\cite{Kadastik:2009dj,Frigerio:2009wf}. A brief introduction to the
intermediate subgroups and $\mathbb{Z}_2$ symmetry will be given in
Sec.~\ref{sec:intdisc}. Possible SO(10) multiplets that contain an
electric and color neutral component for a WIMP DM candidate are summarized in
Sec.~\ref{sec:wimp}. For a group theoretical argument on the
classification of these DM candidates, see
Appendix~\ref{app:DMcandidates}. Among them, those who have a non-zero
hypercharge are severely restricted by the DM direct search
experiments. We consider this class of DM candidates in
Sec.~\ref{sec:ydmbound} and discuss conditions for the DM models to
evade the direct search bound.  

To keep our model concise, in the following discussion, we only consider
SO(10) irreducible representations with dimensions up to 210. 

\subsection{SO(10) GUT and discrete symmetry}
\label{sec:intdisc}

We start by giving a brief description of the ingredients in our
model. In an SO(10) unification theory, a generation of SM fermions
and a right-handed neutrino are embedded in a ${\bf 16}$ chiral
representation, while the SM Higgs boson usually lies in a $\bf 10$
representation. To obtain a realistic Yukawa sector, it is necessary to
take the ${\bf 10}$ to be complex \cite{Bajc:2005zf,
Fukuyama:2004xs}. We will keep this sector unchanged in most of what follows. 
In addition to the SM particles, the $R_1$ and $R_2$ Higgs
representations are added to break SO(10) and $G_\text{int}$,
respectively. The last ingredient of our model is the DM multiplet,
whose lightest component is targeted to be the DM in the Universe. The
stability of the DM is guaranteed by a remnant $\mathbb{Z}_2$ symmetry of the
extra U(1) gauge symmetry of SO(10) as we will discuss soon. Possible
representations for the DM multiplet are determined below. Here, we
assume that only a minimal set of the Higgs and DM multiplets which are
necessary for the symmetry breaking and mass generation of DM lie in the
low-energy regime and other components have masses of the order of the
symmetry breaking scale at which their masses are generated. For
example, among the ${\bf 10}$ representation, only the electroweak doublet components
remains light to break the electroweak symmetry, while the other
components have GUT-scale masses. Also, to obtain the right relic
abundance, the mass of the DM particle is taken to be of order the TeV scale,
while the masses of the other components are either of ${\cal
O}(M_{\text{GUT}})$ or ${\cal O}(M_{\text{int}})$. Such a hierarchical
mass spectrum is obtained with fine-tunings similar to the
doublet-triplet splitting needed for the ${\bf 10}$. In principle, it
may be possible to achieve the splitting of the DM multiplets with a
more elaborate scheme of particle representations as in the Higgs
doublet-triplet separation \cite{2-3, Dimopoulos:1981xm, Inoue:1985cw}.

SO(10) is a rank-five group so it contains an additional U(1) symmetry
besides the SM gauge symmetry. This additional U(1) can be broken into a
$\mathbb{Z}_2$ symmetry by a VEV of an appropriate Higgs
field. If we restrict our attention to representation of dimension ${\bf 210}$ or less, 
the only choice of an irreducible $R_2$ that ensures the $\mathbb{Z}_2$ symmetry is a ${\bf 126}$.\footnote{The next-to-minimal
possibility is  a ${\bf 1728}$. In addition, we note that a $\mathbb{Z}_3$
symmetry can be obtained by a {\bf 672} Higgs field
\cite{DeMontigny:1993gy, Mambrini:2015vna}. } This 
$\mathbb{Z}_2$ symmetry is equivalent to matter parity
$P_M=(-1)^{3(B-L)}$\cite{Farrar:1978xj}, under which the SM fermions are
odd, while the SM Higgs field is even. Thus a fermion (boson) is stable
if it has an even (odd) matter parity.

We list in Table~\ref{tab:intgauge} all possible rank-five subgroups and
corresponding $R_1$ whose VEV breaks SO(10) to the subgroups. Here $D$
denotes the so-called $D$-parity \cite{Kuzmin:1980yp}, that is, a
$\mathbb{Z}_2$ symmetry with respect to the exchange of
SU(2)$_L\leftrightarrow \text{SU}(2)_R$. $D$-parity can be related to an
element of SO(10) \cite{Kuzmin:1980yp} under which a fermion field
transforms into its charge conjugate. In cases where $D$-parity is
not broken by $R_1$, it is subsequently broken by $R_2$ at the intermediate scale
$M_{\text{int}}$. In this work we only consider the subgroups without an explicit
SU(5) factor. Since the DM is necessarily a color singlet, the running of the strong
gauge coupling is unaltered by the presence of a new DM particle
below the intermediate scale. Thus even though the addition of a DM
multiplet yields unification of the gauge couplings, the unification scale
$M_{\text{int}}$ is always less than $10^{14}$~GeV as the contribution
to the U(1)$_Y$ beta function is always positive. If we now 
associate $M_{\text{int}}$ with SU(5), this low partial unification is heavily disfavored on the
basis of proton lifetime constraints.
Flipped SU(5) usually has a high
intermediate scale and a high GUT scale close to the
Planck scale. In this case higher dimension operators suppressed by
Planck scale become important, and one may also need to rely on a double
seesaw for the explanation of neutrino masses
\cite{Antoniadis:1987dx, flipped}. These bring extra complication into
our model and we do not consider these possibilities here. Other intermediate
gauge groups in the table are subgroups of $\text{SU}(4)_C\otimes
\text{SU}(2)_L \otimes \text{SU}(2)_R\otimes {D}$, and
$\text{U(1)}_{B-L}$ is a subgroup of $\text{SU}(4)_C$. The relationship
among hypercharge $Y$, the U(1)$_{B-L}$ charge $B-L$, and the third
component of the SU(2)$_R$ generators $T^3_R$ is very useful for
determining the quantum numbers of DM candidates: 
\begin{equation}
Y=\frac{B-L}{2}+T^3_R~.
\end{equation}
The convention we are using for hypercharge is such that electric charge is given by
$Q=T^3_L+Y$, with $T^3_L$ denoting the third component of the SU(2)$_L$
generators. 

\begin{table}[ht!]
 \begin{center}
\caption{\it Candidates for the intermediate gauge group $G_{\text{int}}$.}
\label{tab:intgauge}
\vspace{5pt}
\begin{tabular}{ll}
\hline
\hline
$G_{\text{int}}$ & $R_1$ \\
\hline
$\text{SU}(4)_C\otimes \text{SU}(2)_L \otimes \text{SU}(2)_R$& {\bf 210}\\
$\text{SU}(4)_C\otimes \text{SU}(2)_L \otimes \text{SU}(2)_R\otimes {D}$& {\bf
     54}\\
$\text{SU}(4)_C\otimes \text{SU}(2)_L \otimes \text{U}(1)_R$ & {\bf 45}\\
$\text{SU(3)}_C\otimes \text{SU}(2)_L \otimes \text{SU}(2)_R
 \otimes \text{U}(1)_{B-L}$ &{\bf 45}\\
$\text{SU(3)}_C\otimes \text{SU}(2)_L \otimes \text{SU}(2)_R
 \otimes \text{U}(1)_{B-L} \otimes D$ & {\bf 210}\\
$\text{SU(3)}_C\otimes \text{SU}(2)_L \otimes \text{U}(1)_R 
 \otimes \text{U}(1)_{B-L}$ & {\bf 45}, {\bf 210}\\
\hline
$\text{SU}(5) \otimes \text{U}(1)$ & {\bf 45}, {\bf 210}\\
$\text{Flipped}~ \text{SU}(5) \otimes \text{U}(1)$ & {\bf 45}, {\bf 210}\\
\hline
\hline
\end{tabular}
 \end{center}
\end{table}

To summarize, our model contains the usual SM content and an
$\text{SU(2)}_L$ multiplet for DM at low energy scale. At the
intermediate scale, depending on $G_\text{int}$ there are parts of the DM
SO(10) multiplet, parts of the ${\bf 126}$ Higgs field to break
$G_\text{int}$ while conserving matter parity, and perhaps some other Higgs
fields that we specify on a model by model basis and are necessary for fine-tuning the DM mass. All other
components in SO(10) multiplets are assumed to be at the GUT scale.

\subsection{WIMP DM candidates}
\label{sec:wimp}

In this section we discuss possible DM candidates in SO(10)
representations up to $\bf 210$, and classify them according to their
quantum numbers. As discussed in the last section, the stability of DM is
ensured by matter parity. Thus a fermionic DM candidate should be
parity even and belong to a ${\bf 10}$, ${\bf 45}$, ${\bf 54}$,
${\bf 120}$, ${\bf 126}$, ${\bf 210}$ or ${\bf 210}^{\prime}$
representation, while scalar DM is 
parity odd and belongs to a $\bf 16$ or $\bf 144$
representation \cite{Martin:1992mq, DeMontigny:1993gy,
Mambrini:2015vna}. Following the branching rules given in
Ref.~\cite{Slansky:1981yr}, in Table~\ref{tab:dmcandidate}, we list
$\text{SU(2)}_L\otimes\text{U(1)}_Y$ multiplets in various SO(10)
representations that contain an electrically neutral color singlet. A similar
list of candidates can be found in earlier work
\cite{Frigerio:2009wf}. The table is classified by $B-L$ so one can
check the matter parity of the candidates easily; $B-L=0,~2$ candidates
are fermionic while $B-L=1$ candidates are scalar, labeled by an ``{\tt
F}'' or ``{\tt S}'' at the beginning of each row, respectively. The
subscript of the model names denotes the $\text{SU}(2)_L$
representation, while the superscript shows hypercharge. A hat is used for
$B-L = 2$ candidates.

\begin{table}[t]
 \begin{center}
\caption{\it List of $\text{SU(2)}_L\otimes\text{U(1)}_Y$ multiplets in
  SO(10) representations that contain an electric neutral color singlet. }
\label{tab:dmcandidate}
\vspace{5pt}
\begin{tabular}{l c c l l}
\hline
\hline
 Model & $B-L$ & $\text{SU(2)}_L$ & $Y\qquad$ &  SO(10) representations\\
\hline
\model{F}{1}{0} &\multirow{6}{*}{0}
 & {\bf 1} & $0$ &   {\bf 45}, {\bf 54}, {\bf 210} \\
\model{F}{2}{1/2} & & {\bf 2} & $1/2$ & {\bf 10}, {\bf 120}, {\bf 126}, ${\bf 210}^\prime$ \\
\model{F}{3}{0}& & {\bf 3} & $0$ &  {\bf 45}, {\bf 54}, {\bf 210} \\ 
\model{F}{3}{1}& & {\bf 3} & $1$ &   {\bf 54} \\ 
\model{F}{4}{1/2}& & {\bf 4} & $1/2$ &  ${\bf 210}^\prime$ \\
\model{F}{4}{3/2}& & {\bf 4} & $3/2$ &  ${\bf 210}^\prime$ \\
\hline
\model{S}{1}{0} & \multirow{4}{*}{1}
 & {\bf 1} & $0$ &   {\bf 16}, {\bf 144} \\
\model{S}{2}{1/2}& & {\bf 2} & $1/2$&    {\bf 16}, {\bf 144} \\
\model{S}{3}{0}& & {\bf 3} & $0$&  {\bf 144} \\
\model{S}{3}{1}& & {\bf 3} & $1$&  {\bf 144} \\ 
\hline
\model{\widehat{F}}{1}{0}& \multirow{3}{*}{2}
 & {\bf 1} & $0$ &   {\bf 126} \\
\model{\widehat{F}}{2}{1/2}& & {\bf 2} & $1/2$&   {\bf 210} \\
\model{\widehat{F}}{3}{1}& & {\bf 3} & $1$& {\bf 126} \\
\hline
\hline
\end{tabular}
 \end{center}
\end{table}

We consider the WIMP DM scenario, which requires DM to be in thermal
equilibrium with the SM particles before its abundance freezes out. This
generally requires DM particles to interact with SM particles
efficiently in the early universe. As a consequence, the fermionic
singlets \model{F}{1}{0} and \model{\widehat{F}}{1}{0} are not good WIMP
candidates since they are SM 
singlets and can only interact with SM particles through exchange of
intermediate scale virtual particles. In fact, these are examples of NETDM
and the possibilities for \model{F}{1}{0} and \model{\widehat{F}}{1}{0}
candidates for DM were discussed extensively in
\cite{Mambrini:2015vna}. Indeed, there it was shown that only two NETDM
candidates from SO(10) survive all phenomenological constraints. One
possibility is associated with the $\text{SU}(4)_C \otimes
\text{SU}(2)_L \otimes \text{SU}(2)_R$ intermediate gauge group. In this
case, the DM candidate is in a  ($\bf{1,1,3}$) originating in a ${\bf
45}$. This is an example of \model{F}{1}{0}. The second example is based
on $\text{SU}(4)_C \otimes \text{SU}(2)_L \otimes \text{SU}(2)_R \otimes
D$ and consists of a $(\bf{15,1,1}$) originating from either a ${\bf 45}$ or a ${\bf 210}$
in SO(10). Since the ${\bf 15}$ of SU(4)$_C$ carries zero $B-L$ charge,
this is also an example of \model{F}{1}{0}. All possible candidates
associated with \model{\widehat{F}}{1}{0} were excluded in
\cite{Mambrini:2015vna}. A fermion that is a singlet under the
intermediate gauge group can also be produced through the exchange of
the GUT scale particles, and thus be a DM candidate. For example, the
case of the $({\bf 1}, {\bf 1}, {\bf 1})$ component of a ${\bf 210}$ is
discussed in Ref.~\cite{Mambrini:2015vna}, which is again an example of
\model{F}{1}{0} DM.

The scalar singlet \model{S}{1}{0} and triplet \model{S}{3}{0} can
interact with the SM Higgs boson efficiently through the quartic
coupling and are potential good DM candidates to be discussed
below. These can be taken to be either real or complex. For \model{S}{1}{0},
there is no difference in any of our results whether \model{S}{1}{0} is real or complex. We
have taken \model{S}{3}{0} to be real, but there would be no qualitative difference in our
results for complex \model{S}{3}{0}. 
In addition,  \model{S}{3}{0} couples to the SM particles via the weak
interaction. Similarly, the fermion triplet \model{F}{3}{0} is a
wino-like DM candidate and will also be considered below. In general, the neutral
component of a $\text{SU(2)}_L\otimes\text{U(1)}_Y$ multiplet can interact with
SM particles through exchange of $W$ or $Z$ boson, and thus can be a
good DM candidate. Such DM candidates
have been widely studied in the literature~\cite{Cirelli:2005uq,
Essig:2007az, Hambye:2009pw, Hisano:2014kua,
Nagata:2014wma, Nagata:2014aoa, Boucenna:2015haa,
Harigaya:2015yaa,Heeck:2015qra, Cirelli:2015bda,Chiang:2015fta}. 

There are also DM candidates which have non-zero hypercharge. 
These are: \model{F}{2}{1/2}, \model{F}{3}{1},
\model{F}{4}{1/2}, \model{F}{4}{3/2}, \model{S}{2}{1/2},
\model{S}{3}{1}, \model{\widehat{F}}{2}{1/2}, and
\model{\widehat{F}}{3}{1}. These DM candidates are
severely constrained by DM direct detection experiments since their
scattering cross sections with a nucleon induced by $Z$-boson exchange
are generally too large. Possible ways to evade this constraint are discussed in
the following section.

\subsection{Hypercharged DM}
\label{sec:ydmbound}

A DM candidate with $Y\neq 0$ needs to be a Dirac spinor or a complex
scalar, depending on its matter parity. These hypercharged candidates
are severely restricted by the direct detection experiments, since they
elastically scatter nucleons via the vector interactions mediated by
$Z$-boson exchange, whose scattering cross section turns out to be too
large by many orders of magnitude. 
One possible way to evade the constraint is to generate mass splitting,
$\Delta m$, between the neutral components of such a DM multiplet $\psi$
and to split it into two Majorana fermions or real scalars $\chi_1$,
$\chi_2$. Then, the DM no longer suffers from large scattering cross
sections since it does not have vector interactions. Such
splitting occurs if the DM mixes with extra
$\text{SU(2)}_L\otimes \text{U(1)}_Y$ multiplets after electroweak
symmetry breaking, just like higgsinos in the MSSM, which originally
form a Dirac fermion, reduce to neutralinos after they mix with gauginos. 
As we have assumed that only a single DM multiplet lies in the low-energy region,
a natural mass scale for the extra $\text{SU(2)}_L\otimes \text{U(1)}_Y$
multiplets is the intermediate scale $M_{\text{int}}$. The effects of
these heavy particles on the low-energy theory are expressed in terms of
effective operators induced after integrating them out. Among them, the
following operator is relevant for the generation of mass splitting for
Dirac fermion DM:
\begin{equation}
\frac{1}{M^n_\text{int}}\overline{\psi^{\cal C}}\psi H^{*p}~,
\label{eq:effmass}
\end{equation}
where $H$ is the SM Higgs field, ${\cal C}$ represents charge conjugation,
$p=4Y_{\psi}$ with $Y_\psi>0$ being the hypercharge of the DM $\psi$, and
$n=p-1$. For complex scalar DM, we 
have a similar operator with $n=p-2$. Notice that the above operator
violates any particle number assigned to the fermion $\psi$. After the Higgs
field acquires a VEV, the above operator reduces to a Majorana mass
term, which generates a mass splitting between two Weyl fermions inside
the neutral component of the DM multiplet. Namely, the mass eigenstates are expressed by two Majorana
fermions, and the lighter one can be regarded as DM. The splitting is
just given  by $\Delta m \sim v^p/M^n_{\text{int}}$ with $v\simeq
174$~GeV being the Higgs VEV. Since a Majorana
fermion does not couple to vector interactions, DM-nucleon elastic
scattering cross sections are significantly reduced and we can avoid the
direct detection bound. In the case of scalars, 
a similar operator to that in Eq.~\eqref{eq:effmass} induces a splitting in
the squared masses and hence $\Delta m \sim v^p/(M^n_{\text{int}}
m_\psi)$ where $m_\psi$ corresponds to the scalar mass term $m_\psi^2
\psi \psi^*$.

However, notice that the operator \eqref{eq:effmass} is considerably
suppressed if the intermediate scale is large \cite{Nagata:2014aoa}. The
suppression becomes more significant when $Y_\psi$ is larger. In this
case, the resultant mass splitting becomes extremely small. When
the mass splitting is sufficiently small, then the DM can scatter off a
nucleon $N$ inelastically: $\chi_1+N\rightarrow \chi_2+N$. Since this
process is induced by the vector interactions, the scattering cross section
again becomes too large if the mass splitting $\Delta m$ is smaller than
the recoil energy. This sets the bound $\Delta m\gtrsim
100~\text{keV}$. For this condition to be satisfied, $M_\text{int}\lesssim
10^9$, $3\times 10^4$, and $4\times 10^3$~GeV are required for fermionic
dark DM with $Y_\psi=1/2$, 1 and 3/2, respectively
\cite{Nagata:2014aoa}. In the case of scalar DM, the upper bound depends
on the DM mass. For a 1~TeV DM mass, $M_{\text{int}} \lesssim 10^5$~GeV
for $Y_\psi = 1$. For a $Y_\psi=1/2$ scalar DM candidate, on the other hand,
the mass splitting can be generated with a renormalizable interaction and
its effect on the mass splitting depends only on its dimensionless
coefficient. We will see later that this coefficient can still be very
small, whose size is determined by the symmetry breaking pattern and its
scale. This is because the operators relevant for the generation of the
mass splitting are forbidden by the SO(10) symmetry. Thus, the
constraint from inelastic scattering can again give a bound on the
intermediate scale even for $Y_\psi=1/2$ scalar DM candidates. 

Another possibility to evade the direct detection bound is to push the
DM mass sufficiently high. Since the local number density of DM is inversely
proportional to the DM mass as the DM local energy density is fixed, the
DM direct detection constraints are relaxed if the DM mass is taken to
be heavy enough. In this case, DM is produced non-thermally
\cite{Feldstein:2013uha}. This possibility is also discussed below.


\section{Scalar dark matter}
\label{sec:scalarDM}

In this section, we discuss scalar WIMP DM in SO(10) models with
different intermediate subgroups. In this case, the DM candidates belong
to either a $\bf 16$ or a $\bf 144$ representation. The masses
of components in a DM multiplet in general need to be fine-tuned; if a
charged component is nearly degenerate with the DM particle and
decays to it only through an intermediate-scale gauge boson or Higgs
field, this charged particle would be very long lived, which is
cosmologically disastrous \cite{Mambrini:2015vna}. Thus, to be safe, we take
the masses of these extra 
components to be ${\cal O}(M_{\text{GUT}})$ or ${\cal O}(M_{\text{int}})$,
while the DM mass to be around TeV scale so that the thermal relic
abundance of the DM agrees with the observed DM density, as we will see in
Sec.~\ref{sec:scalarDMmass}. Here, we assume
that the fine-tuning of DM masses be realized with a minimal choice of
Higgs fields, that is, we exploit only $R_1$ and $R_2={\bf 126}$ to
generate desired mass spectrum with $R_1$ being an irreducible
representation chosen from Table~\ref{tab:intgauge}. This is possible
because a ${\bf 16}$ or ${\bf 144}$ can couple to the ${\bf 126}$ Higgs
field. Then, we study whether each set of matter content and its mass
spectrum offers gauge coupling unification with appropriate GUT and
intermediate scales. In Sec.~\ref{sec:scalarDM_running}, we present the
results for the analysis and list promising candidates with
$M_\text{int}$ and $M_\text{GUT}$ determined by means of renormalization
group equations (RGEs). The fine-tuning for the masses of components
in a DM multiplet is discussed in
Sec.~\ref{sec:scalarDM_finetuning}. As discussed in the previous
section, hypercharged DM candidates require additional consideration for
the generation of the mass splitting between the neutral components to
avoid the bound from the direct detection experiments. This is discussed
in Sec.~\ref{sec:scalarspliting}. Finally, in
Sec.~\ref{sec:scalarpheno}, we summarize the current experimental
constraints and future prospects for the scalar DM candidates discussed
in this section.

\subsection{DM mass}
\label{sec:scalarDMmass}

To determine the renormalization group (RG) running of gauge couplings, we
need to know the mass of DM candidates, since they affect the
running above its mass scale. An exception is \model{S}{1}{0} as it is a
SM singlet and does not contribute to the gauge coupling beta functions
below $M_\text{int}$ at the one-loop level. Scalar singlet DM is
discussed in Ref.~\cite{Burgess:2000yq}. To roughly estimate favored
mass region for such a singlet DM particle, consider the quartic
interaction between the singlet DM $\phi$ and the SM Higgs field:
$-\lambda_{H\phi} \phi^2 |H|^2/2$. Through this coupling, the singlet DM
particles annihilate into a pair of the SM Higgs bosons. The
annihilation cross section $\sigma_{\text{ann}}$ times the relative
velocity between the initial state particles $v_{\text{rel}}$ is
evaluated as 
\begin{equation}
\sigma_{\text{ann}}v_{\text{rel}} \simeq
\frac{\lambda^2_{H\phi}}{16 \pi m_{\text{DM}}^2}~,
\end{equation} 
assuming that the DM mass $m_{\text{DM}}$ is much larger than the
SM Higgs mass $m_h$ and we neglect terms proportional to 
$v^2$. The DM relic abundance is, on the other hand,
related to the annihilation cross section by
\begin{equation}
\Omega_{\text{DM}} h^2 \simeq  
\frac{3 \times 10^{-27}~\text{cm}^3 ~\text{s}^{-1}}{\langle
 \sigma_{\text{ann}} v_{\text{rel}} \rangle} ~.
\label{eq:DMab}
\end{equation}
To account for the observed DM density $\Omega_{\text{DM}} h^2 = 0.12$~\cite{Ade:2015xua},
the DM mass should be  $m_{\text{DM}}\lesssim 10~\text{TeV}$ for
$\lambda_{H\phi}\lesssim 1$. This gives us a rough upper bound for the DM mass.   

The other scalar DM candidates are $\text{SU(2)}_L\otimes \text{U}(1)_Y$
multiplets, which can interact with SM particles through gauge
interactions besides the quartic coupling mentioned above. In particular,
\model{S}{2}{1/2} is known as the Inert Higgs Doublet Model and has been widely
studied in the literature\footnote{For another approach to the
realization of the Inert Higgs doublet model based on grand unification,
see Ref.~\cite{ky}.}  
\cite{Deshpande:1977rw, Arhrib:2013ela}. 
To evaluate the effects of gauge interactions, let us
first consider the limit of zero quartic couplings. In this case, the
annihilation cross sections are completely determined as functions of
the DM mass $m_{\text{DM}}$. Since the annihilation into SM fermions and
Higgs boson suffers from $p$-wave suppression, the DM particles
annihilate predominantly into the weak gauge bosons. In addition, we need
to take into account coannihilation effects since all of the
components in a $\text{SU(2)}_L$ multiplet are degenerate in mass; the
mass difference among the components is induced after electroweak
symmetry breaking at the loop level and thus is quite suppressed compared to
the DM mass, as small as ${\cal O}(100)$~MeV. Taking these effects into
account, for \model{S}{n}{Y}, the effective (averaged)
annihilation cross section is given by \cite{Cirelli:2005uq} 
\begin{equation}
\sigma_{\text{ann}} v_{\text{rel}} \simeq 
\frac{g^4 (3-4 n^2+n^4) + 16 Y^4 g^{\prime 4}+ 8 g^2 g^{\prime 2}Y^2
(n^2-1)}{64 \pi c_n m_{\text{DM}}^2} ~,
\label{eq:effannscalar}
\end{equation}
where $g$ ($g^{\prime}$) are the $\text{SU(2)}_L$ ($\text{U(1)}_Y$) gauge couplings 
and $c_n = n $ $(2n)$ for a real (complex) scalar. Here, we assume the
DM mass to be much larger than the weak gauge boson masses. Again
Eq.~\eqref{eq:DMab} tells us that the masses of the DM candidates should
fall into a region from $\sim 500$~GeV to $\sim 2$~TeV. On the other
hand, if the quartic coupling is larger than the gauge couplings, the
annihilation into a pair of Higgs bosons becomes dominant and thus the DM
abundance would be similar to that of the singlet DM candidate. In general, the DM
mass should lie between $0.5~\text{TeV}$ to $10~\text{TeV}$ for
\model{S}{2}{1/2}, \model{S}{3}{0} and \model{S}{3}{1}.  

More accurate estimations for the DM mass can be found in the
literature \cite{Cirelli:2005uq, Hambye:2009pw, Farina:2013mla} with various additional contributions taken into
account. For $\text{SU(2)}_L\otimes \text{U}(1)_Y$ DM
multiplets, the non-perturbative Sommerfeld enhancement
is of great importance \cite{Hisano:2003ec}. In the limit of
zero quartic coupling, the DM masses with which the relic abundance
agrees with the observed DM density are evaluated as
$m_\text{DM}=0.5$ and $2.5$~TeV for \model{S}{2}{1/2} and
\model{S}{3}{0}, respectively \cite{Farina:2013mla}. For
\model{S}{3}{1}, as far as we know, there has been no calculation which
includes the Sommerfeld enhancement; thus its mass would be larger than
1.6~TeV, which is obtained with only the perturbative contribution
considered \cite{Cirelli:2005uq}. For the cases where the scalar DM
multiplets have non-zero quartic coupling with the SM Higgs doublet, it
was shown in Ref.~\cite{Hambye:2009pw} that the allowed DM mass can be
extended to $\sim 58$ and $28$~TeV for \model{S}{2}{1/2} and
\model{S}{3}{0}, respectively, when the quartic coupling $\lambda \sim 4
\pi$. Such a large quartic coupling is, however, in general inconsistent
with GUTs since it immediately blows up at a scale much below the GUT
and intermediate scales. Thus, we implicitly assume the quartic coupling
should be rather small, \textit{e.g.}, $\lesssim 1$, to avoid divergent
couplings. In this case, the DM mass usually lies around ${\cal O}(1)$~TeV.

\subsection{Candidates for scalar DM}
\label{sec:scalarDM_running}

\begin{table}[t!]
\centering
\caption{\it Summary of DM multiplets. The second column shows the
 $G_\text{int}$  representation with quantum numbers listed in the same
 order as the groups shown in the direct product. The case of $G_{\rm
 int}=\text{SU}(4)_C\otimes \text{SU}(2)_L\otimes \text{SU}(2)_R\otimes
 D$ ($\text{SU}(3)_C\otimes \text{SU}(2)_L\otimes \text{SU}(2)_R \otimes
 \text{U}(1)_{B-L}\otimes D$) is identical to that of $G_{\rm
 int}=\text{SU}(4)_C\otimes \text{SU}(2)_L\otimes \text{SU}(2)_R$
 ($\text{SU}(3)_C\otimes \text{SU}(2)_L\otimes \text{SU}(2)_R \otimes
 \text{U}(1)_{B-L}$) with additional multiplets required by left-right
 symmetry introduced above the intermediate scale.  }
\label{table:scalar_model}
\vspace{5pt}
  \begin{tabular}{lllc}
    \hline
    \hline
   Model~~   & $R_\text{DM}$  & \model{S}{n}{Y} & SO(10) representation\\ 
    \hline
    \hline
    \multicolumn{4}{c}{$G_{\rm int}=\text{SU}(4)_C\otimes
   \text{SU}(2)_L\otimes \text{SU}(2)_R (\otimes D)$}\\
    \hline
\DM{SA}{422(D)} & ${\bf 4},{\bf 1},{\bf 2}$ & \model{S}{1}{0} & ${\bf
	       16}$, ${\bf 144}$ \\  
   \DM{SB}{422(D)} & ${\bf 4},{\bf 2},{\bf 1}$ & \model{S}{2}{1/2} &
	       ${\bf 16}$, ${\bf 144}$\\  
   \DM{SC}{422(D)} & ${\bf 4},{\bf 2},{\bf 3}$ & \model{S}{2}{1/2} &
	       ${\bf 144}$\\   
   \DM{SD}{422(D)} & ${\bf 4},{\bf 3},{\bf 2}$ &  \model{S}{3}{1} &
	       ${\bf 144}$\\  
   \DM{SE}{422(D)} & ${\bf 4},{\bf 3},{\bf 2}$ &  \model{S}{3}{0} &
	       ${\bf 144}$ \\  
   \bhline{1.5pt}
    \multicolumn{4}{c}{$G_{\rm int}=\text{SU}(4)_C\otimes
   \text{SU}(2)_L\otimes \text{U}(1)_R$}\\
 \hline
   \DM{SA}{421} & ${\bf 4},{\bf 1},{-1/2}$~~ & \model{S}{1}{0} & ${\bf
	       16}$, ${\bf 144}$ \\   
 \DM{SB}{421} & ${\bf 4},{\bf 2},{ 0}$ &  \model{S}{2}{1/2} & ${\bf
	       16}$, ${\bf 144}$\\  
   \DM{SC}{421} & ${\bf 4},{\bf 2},{ 1}$ & \model{S}{2}{1/2} & ${\bf 144}$\\  
   \DM{SD}{421} & ${\bf 4},{\bf 3},{ 1/2}$ &  \model{S}{3}{1} & ${\bf 144}$\\ 
   \DM{SE}{421} & ${\bf 4},{\bf 3},{-1/2}$ &  \model{S}{3}{0} & ${\bf
	       144}$ \\    
   \bhline{1.5pt}
    \multicolumn{4}{c}{$G_{\rm int}=\text{SU}(3)_C\otimes
   \text{SU}(2)_L\otimes \text{SU}(2)_R \otimes \text{U}(1)_{B-L}
   (\otimes D)$}\\ 
    \hline
   \DM{SA}{3221(D)}~ & ${\bf 1},{\bf 1},{\bf 2},{ 1}$ & \model{S}{1}{0}
	   & ${\bf 16}$, ${\bf 144}$ \\  
   \DM{SB}{3221(D)} & ${\bf 1},{\bf 2},{\bf 1},{ -1}$ &
	   \model{S}{2}{1/2} & ${\bf 16}$, ${\bf 144}$ \\   
   \DM{SC}{3221(D)} & ${\bf 1},{\bf 2},{\bf 3},{ -1}$ &
	   \model{S}{2}{1/2} & ${\bf 144}$ \\  
   \DM{SD}{3221(D)} & ${\bf 1},{\bf 3},{\bf 2},{ 1}$ & \model{S}{3}{1} &
	       ${\bf 144}$ \\  
   \DM{SE}{3221(D)} & ${\bf 1},{\bf 3},{\bf 2},{ 1}$ &  \model{S}{3}{0}
	   & ${\bf 144}$ \\  
 \hline 
 \hline
  \end{tabular}
\end{table}

We list all possible scalar DM candidates in
Table~\ref{table:scalar_model}. All of the candidates belong to either
a ${\bf 16}$ or a ${\bf 144}$. Here, the first column shows the model
names with subscript representing the intermediate gauge group
$G_\text{int}$. The second column lists the $G_\text{int}$
representations that contain the DM candidate multiplet
\model{S}{n}{Y}. All of the components in the representation except the
DM multiplet \model{S}{n}{Y} shown in the third column will have masses tuned to
$\mathcal{O}(M_\text{int})$. The rest of the components in the SO(10)
multiplet have masses of ${\cal O}(M_\text{GUT})$. The case of
$G_{\rm int}=\text{SU}(4)_C\otimes \text{SU}(2)_L\otimes
\text{SU}(2)_R\otimes D$ ($\text{SU}(3)_C\otimes \text{SU}(2)_L\otimes
\text{SU}(2)_R \otimes \text{U}(1)_{B-L}\otimes D$) is identical to
that of $G_{\rm int}=\text{SU}(4)_C\otimes \text{SU}(2)_L\otimes
\text{SU}(2)_R$ ($\text{SU}(3)_C\otimes \text{SU}(2)_L\otimes
\text{SU}(2)_R \otimes \text{U}(1)_{B-L}$) with additional multiplets
required by the left-right symmetry introduced above the intermediate scale.

Consider, for example, \DM{SA}{422}; the $({\bf 4},{\bf 1},{\bf 2})$ DM
multiplet originating from a ${\bf 16}$ of SO(10) in the
$\text{SU}(4)_C\otimes \text{SU}(2)_L\otimes \text{SU}(2)_R$ model. The
``other half'' of the ${\bf 16}$, $({\bf 4},{\bf 2},{\bf 1})$, will have
a GUT scale mass, while 7 of 8 fields in $({\bf 4},{\bf 1},{\bf 2})$ are
tuned to have an intermediate scale mass and only the DM singlet is
tuned to have a weak scale mass.  In the case of \DM{SA}{422D}, the DM multiplet
corresponds to the $({\bf 4},{\bf 1},{\bf 2})\oplus(\overline{\bf 4},{\bf
2},{\bf 1})$ in the $\text{SU}(4)_C\otimes \text{SU}(2)_L\otimes
\text{SU}(2)_R\otimes D$ model. In this case, none of the components have GUT scale masses
and 15 of the 16 fields have intermediate scale masses. 
Thus the presence of the left-right symmetry affects only the field content above the 
intermediate scale, though this will ultimately affect the scale of gauge coupling unification.
These representations are added to a
minimal SO(10) unification model containing three generations of $\bf
16$ chiral multiplet, a complex $\bf 10$ Higgs multiplet, a $R_2=\bf
126$ Higgs multiplet, and a $R_1$ Higgs multiplet chosen from
Table~\ref{tab:intgauge}. Notice that \model{S}{2}{1/2} is contained in
both the model \DM{SB}{}'s and \DM{SC}{}'s. The difference between the
models is the SU(2)$_R$ (or additional U(1)) charge assignment; for
instance, \DM{SB}{422} (\DM{SC}{422}) includes the SU(2)$_R$ singlet
(triplet) DM. From Table~\ref{table:scalar_model}, we find that a ${\bf
16}$ contains only \DM{SA}{}'s and \DM{SB}{}'s, while a ${\bf 144}$ has
all of the candidates listed in the table. 

\begin{table}[tbp]
\centering
\caption{
\it One-loop result for scales, unified couplings, and proton lifetimes
 for models in table.~\ref{table:scalar_model}. The DM mass is set to be
 $m_\text{DM}= 1~\text{TeV}$. The mass scales are given in GeV and the
 proton lifetimes are in units of years. Blue shaded models evade the
 proton decay bound, $\tau(p\rightarrow e^+ \pi^0)>1.4\times
10^{34}~\text{yrs}$\cite{Shiozawa, Babu:2013jba}.}
\label{table:scalar_result1} 
\vspace{5pt}
\scalebox{0.93}{
  \begin{tabular}{lllll}
\hline
\hline
 Model    & $\log_{10} M_{\rm GUT}$  & $\log_{10}M_{\rm int}$ &
   $\alpha_{\rm GUT}$  & $\log_{10}\tau_p(p\rightarrow e^+ \pi^0)$ \\ 
    \hline  
 \hline 
    \multicolumn{5}{c}{$G_{\rm int}=\text{SU}(4)_C\otimes
   \text{SU}(2)_L\otimes \text{SU}(2)_R$}\\
 \hline
\rowcolor{LightBlue}
 \DM{SA}{422} & $16.33$ & $11.08$ & $0.0218$~~~~ & $ 36.8 \pm 1.2$ \\ 
\rowcolor{LightBlue}
 \DM{SB}{422} & $15.62$ & $12.38$ & $0.0228$ & $ 34.0 \pm 1.2$ \\ 
 \DM{SC}{422} & $14.89$ & $11.18$ & $0.0243$ & $ 31.0 \pm 1.2$ \\ 
 \DM{SD}{422} & $14.11$ & $13.29$ & $0.0253$ & $ 28.0 \pm 1.2$ \\ 
 \DM{SE}{422} & $14.73$ & $13.72$ & $0.0243$ & $ 30.4 \pm 1.2$ \\ 
    \bhline{1.5pt}
    \multicolumn{5}{c}{$G_{\rm int}=\text{SU}(4)_C\otimes
   \text{SU}(2)_L\otimes \text{SU}(2)_R \otimes D$}\\
\hline
\DM{SA}{422D} & $15.23$ & $13.71$ & $0.0245$ & $ 32.4 \pm 1.2$ \\ 
\DM{SB}{422D} & $15.01$ & $13.71$ & $0.0247$ & $ 31.6 \pm 1.2$ \\  
\DM{SC}{422D} & $14.50$ & $13.71$ & $0.0254$ & $ 29.5 \pm 1.2$ \\ 
\DM{SD}{422D} & $13.95$ & $13.47$ & $0.0260$ & $ 27.3 \pm 1.2$ \\ 
\DM{SE}{422D} & $14.55$ & $13.96$ & $0.0251$ & $ 29.7 \pm 1.2$ \\ 
\bhline{1.5pt}
    \multicolumn{5}{c}{$G_{\rm int}=\text{SU}(4)_C\otimes
   \text{SU}(2)_L\otimes \text{U}(1)_R$}\\
    \hline
\DM{SA}{421} & $14.62$ & $10.96$ & $0.0226$ & $ 30.1 \pm 1.2$ \\ 
\DM{SB}{421} & $14.55$ & $11.90$ & $0.0233$ & $ 29.8 \pm 1.2$ \\ 
\DM{SC}{421} & $14.15$ & $10.92$ & $0.0236$ & $ 28.2 \pm 1.2$ \\ 
\DM{SD}{421} & $13.91$ & $12.80$ & $0.0250$ & $ 27.2 \pm 1.2$ \\ 
\DM{SE}{421} & $14.45$ & $13.12$ & $0.0241$ & $ 29.4 \pm 1.2$ \\ 
\bhline{1.5pt}
    \multicolumn{5}{c}{$G_{\rm int}=\text{SU}(3)_C\otimes
   \text{SU}(2)_L\otimes \text{SU}(2)_R \otimes \text{U}(1)_{B-L}$}\\
    \hline
\rowcolor{LightBlue}
\DM{SA}{3221} & $16.66$ & $ 8.54$ & $0.0217$ & $ 38.1 \pm 1.2$ \\ 
\rowcolor{LightBlue}
\DM{SB}{3221} & $16.17$ & $ 9.80$ & $0.0223$ & $ 36.2 \pm 1.2$ \\ 
\rowcolor{LightBlue}
\DM{SC}{3221} & $15.62$ & $ 9.14$ & $0.0230$ & $ 34.0 \pm 1.2$ \\  
\DM{SD}{3221} & $14.49$ & $12.07$ & $0.0246$ & $ 29.5 \pm 1.2$ \\ 
\DM{SE}{3221} & $15.09$ & $12.22$ & $0.0237$ & $ 31.9 \pm 1.2$ \\ 
\bhline{1.5pt}
     \multicolumn{5}{c}{$G_{\rm int}=\text{SU}(3)_C\otimes
   \text{SU}(2)_L\otimes \text{SU}(2)_R \otimes \text{U}(1)_{B-L}\otimes D$}\\
    \hline
\rowcolor{LightBlue}
\DM{SA}{3221D} & $15.58$ & $10.08$ & $0.0231$ & $ 33.8 \pm 1.2$ \\ 
\rowcolor{LightBlue}
\DM{SB}{3221D} & $15.40$ & $10.44$ & $0.0233$ & $ 33.1 \pm 1.2$ \\   
\DM{SC}{3221D} & $14.58$ & $11.62$ & $0.0245$ & $ 29.8 \pm 1.2$ \\  
\DM{SD}{3221D} & $14.07$ & $12.13$ & $0.0253$ & $ 27.8 \pm 1.2$ \\ 
\DM{SE}{3221D} & $14.60$ & $12.29$ & $0.0245$ & $ 29.9 \pm 1.2$ \\ 
 \hline 
 \hline
  \end{tabular}
}
\end{table}

Next, we perform the RGE\footnote{The beta functions for the minimal
SO(10) GUT described above are given in Appendix B of
Ref.~\cite{Mambrini:2015vna}.} analysis in the models presented in
Table~\ref{table:scalar_model} to see if these models achieve gauge
coupling unification with appropriate GUT and intermediate scales. The
one-loop results for $M_\text{GUT}$, $M_\text{int}$, the unified gauge coupling
$\alpha_\text{GUT}$, and the proton lifetimes in the $p\rightarrow e^+
\pi^0$ channel are shown in Table~\ref{table:scalar_result1}.\footnote{We restrict our attention to one-loop running
as two loop effects become very model dependent on our choice of the scalar potential.} Here,
$M_\text{GUT}$ and $M_\text{int}$ are given in GeV units, while the unit for
proton lifetimes $\tau_p(p\rightarrow e^+ \pi^0)$ is years. 
The DM mass is set to be $m_\text{DM}= 1~\text{TeV}$. We have checked
that altering the DM mass by an order of magnitude results in only
a ${\cal O}(0.2)$\% variation in the logarithmic masses of  $M_\text{int}$ and
$M_\text{GUT}$.  The uncertainty of the lifetime reflects our innocence
of the GUT-scale gauge boson mass $M_X$, which we take it to be within a
range of $0.5M_\text{GUT}\lesssim M_X\lesssim 2M_\text{GUT}$.  It turns
out that most models have already been ruled out by the current
experimental constraint $\tau(p\rightarrow e^+ \pi^0)>1.4\times
10^{34}~\text{yrs}$\cite{Shiozawa, Babu:2013jba}. The models that
possibly survive this constraint are \DM{SA}{422}, \DM{SB}{422},
\DM{SA}{3221}, \DM{SB}{3221}, \DM{SC}{3221}, \DM{SA}{3221D}, and
\DM{SB}{3221D}, which are highlighted in blue shading in the table. In
terms of $\text{SU}(2)_L \otimes \text{U}(1)_Y$ assignments, only
\model{S}{1}{0} and \model{S}{2}{1/2} are found to be viable
candidates. Among them, models \DM{SB}{422}, \DM{SC}{3221},
\DM{SA}{3221D}, and \DM{SB}{3221D} predict proton lifetimes close to the
present limit, and thus can be tested in future proton decay
experiments.

\subsection{Fine-tuning of scalar DM multiplets}
\label{sec:scalarDM_finetuning}

In the previous section, we have reduced the possibilities for
$G_\text{int}$ to the only three gauge groups:
$\text{SU}(4)_C\otimes \text{SU}(2)_L\otimes \text{SU}(2)_R$,
$\text{SU}(3)_C\otimes \text{SU}(2)_L\otimes \text{SU}(2)_R \otimes
\text{U}(1)_{B-L}$, and $\text{SU}(3)_C\otimes \text{SU}(2)_L\otimes
\text{SU}(2)_R \otimes \text{U}(1)_{B-L}\otimes D$. According to
Table~\ref{tab:intgauge}, $R_1={\bf 210}$, ${\bf 45}$, and ${\bf 210}$
yield the above intermediate gauge groups, respectively. In this section,
we briefly discuss how to obtain a desired mass spectrum for the DM multiplet
using these $R_1$'s and $R_2={\bf 126}$ with the help of fine-tuning.
For convenience, we show an explicit procedure for the
fine-tuning in Appendix~\ref{sec:exfinetune}, by taking $R_{\text
DM}={\bf 16}$ and $G_{\rm int}=\text{SU}(3)_C\otimes
\text{SU}(2)_L\otimes \text{SU}(2)_R \otimes \text{U}(1)_{B-L}$ as an
example.  

Let us first write down relevant terms for the mass terms of the DM
multiplet $R_{\text{DM}}$:\footnote{In addition, there are couplings
between the DM and the SM Higgs fields, which give a mass of the
order of the electroweak scale to the DM multiplet. }
\begin{align}
-\mathcal{L}_{\rm int}
&=
M^2 |{R}_{\rm DM}|^2
+ \kappa_1 {R}_{\rm DM}^* R_{\rm DM} R_1
+\{\kappa_2 R_{\rm DM}R_{\rm DM}{R}^*_2 +\text{h.c.}\}
\nonumber \\[3pt]
& + \lambda^{\bf 1}_1 |{R}_{\rm DM}|^2|{R}_{1}|^2
+ \lambda^{\bf 1}_2 |{R}_{\rm DM}|^2|{R}_{2}|^2
+\left\{
\lambda_{12}^{\bf 126}\left( R_{\rm DM}R_{\rm DM}\right)_{\bf 126}
\left(R_1 {R}_2^*\right)_{\overline{\bf 126}}
+\text{h.c.}\right\}
 \nonumber \\[3pt]
&+ \lambda^{\bf 45}_1
\left({R}^*_{\rm DM}R_{\rm DM}\right)_{\bf
 45}\left({R}^*_1R_1\right)_{\bf 45}  
 + \lambda^{\bf 210}_1\left({R}^*_{\rm DM}R_{\rm DM}\right)_{\bf 210}
\left({R}_1^*R_1\right)_{\bf 210}
 \nonumber \\[3pt]
&+ \lambda^{\bf 45}_2
\left({R}^*_{\rm DM}R_{\rm DM}\right)_{\bf
 45}\left({R}^*_2R_2\right)_{\bf 45}  
 + \lambda^{\bf 210}_2\left({R}^*_{\rm DM}R_{\rm DM}\right)_{\bf 210}
\left({R}^*_2R_2\right)_{\bf 210}
~,
\label{eq:lintsc}
\end{align}
where the subscripts after the parentheses denote the SO(10)
representation formed by the product in them. $M$, $\kappa_1$, and
$\kappa_2$ are dimensionful parameters, which we assume to be ${\cal
O}(M_\text{GUT})$. Notice that the term $\left( R_{\rm
DM}R_{\rm DM}\right)_{\bf 120} \left(R_1 {R}_2^*\right)_{{\bf 120}}$ and
its charge conjugate vanish since the $R_{\text{DM}}$ is a bosonic field
and $(AB)_{\bf 120}$ is anti-symmetric with respect to the exchange of
$A$ and $B$. In addition, the term $\left( R_{\rm
DM}R_{\rm DM}\right)_{\bf 10} \left(R_1 {R}_2^*\right)_{{\bf 10}}$ does
not give a mass term for $R_{\text{DM}}$; $\langle R_1 {R}_2^* \rangle$
is singlet with respect to the SM gauge interactions, and a {\bf 10}
representation does not contain such a component. 
The terms with the coefficients $\lambda_1^{\bf 1}$ and
$\lambda_2^{\bf 1}$ are irrelevant to the generation of the mass
splitting in the DM multiplet, as they only give a common mass to all of
the components in the multiplet. 
It is also worth noting that terms including $\kappa_2$
and $\lambda_{12}^{\bf 126}$ break the particle
number which can be assigned to the complex scalar
$R_{\text{DM}}$. Hence, these effects can split $R_{\text{DM}}$
into two real scalars with different masses. We use these interactions
to avoid the direct detection bound in the case of the complex
hypercharged DM, models \DM{SB}{}'s and \DM{SC}{}'s, which we discuss
in the following section. 

After $R_1$ gets a VEV, the terms with $\kappa_1$, $\lambda_1^{\bf 45}$,
and $\lambda_1^{\bf 210}$ generate mass terms for the components in
$R_{\text{DM}}$ with different mass values, since the $R_1$ VEV couples
to them with different Clebsch-Gordan coefficients. Thus, by fine tuning
the coefficients $M$, $\kappa_1$, $\lambda_1^{\bf 45}$, and
$\lambda_1^{\bf 210}$, one can arrange that the DM multiplet obtains a mass of
$\mathcal{O}(M_\text{int})$, with other multiplets remaining around
$\mathcal{O}(M_\text{GUT})$.   

The next step is to separate the $\text{SU(2)}_L$ multiplet
\model{S}{n}{Y} from the intermediate gauge group multiplet. 
This can be accomplished by appropriately tuning the coefficients of
$\kappa_2$, $\lambda_{12}^{\bf 126}$, $\lambda_2^{\bf 45}$, and
$\lambda_2^{\bf 210}$ so that the generated mass terms cancel out the
intermediate scale mass obtained previously, leaving only the DM
candidate at TeV scale.\footnote{In the cases of models \DM{SB}{}
and {\tt SC}, however, terms with $\kappa_2$ and $\lambda_{12}^{\bf
126}$ do not give mass terms for the DM multiplet after the intermediate
symmetry breaking. The reason is as follows. For {\tt SB}, since the
$R_2$ VEV has a SU(2)$_R$ charge while ${\tt SB}$ DM candidates do
not, the couplings between the $R_2$ VEV and the ${\tt SB}$ DM are
forbidden by the SU(2)$_R$ symmetry. For {\tt SC}'s, 
since both the DM candidates and the $R_2$ VEV are in SU(2)$_R$ triplets,
a pair of the DM fields should be combined anti-symmetrically to be
coupled to the $R_2$ VEV, but this vanishes because the DM is bosonic. 
In these cases, therefore, we carry out the fine-tuning for
the DM mass only with the coefficients $\lambda_2^{\bf 45}$ and
$\lambda_2^{\bf 210}$. 
\label{ftnt:masssp}} 
After this step, we obtain a mass spectrum in which only the DM
candidate lies around the TeV scale, while its partner fields with
respect to the intermediate gauge symmetry are at $M_{\text{int}}$. The
rest of the components of $R_{\text{DM}}$ have masses of ${\cal
O}(M_{\text{GUT}})$.  For an explicit example, see Appendix B.

\subsection{Mass splitting of hypercharged scalar dark matter}
\label{sec:scalarspliting}

As discussed in Sec.~\ref{sec:ydmbound}, we need a mass splitting of
$\Delta m\gtrsim 100 {\rm keV}$\cite{Nagata:2014aoa} between the neutral and charged
components of the hypercharged DM candidates (models \DM{SB}{} and
\DM{SC}{}) to avoid the direct detection bound. Since both of these
models yield \model{S}{2}{1/2} DM, the mass splitting can be induced by
dimension-four operators like $\phi^2H^{*2}$, where $\phi$ denotes the
hypercharged scalar DM \model{S}{2}{1/2}. Such operators are, however,
forbidden by the SO(10) GUT symmetry; in fact, as the \model{S}{2}{1/2}
DM and the SM Higgs field have $B-L = 1$ and $0$, respectively, the
operators contributing the mass splitting violate the $B-L$
symmetry. Thus, they can be induced only below the intermediate scale
where the $B-L$ symmetry is spontaneously broken. 

\begin{figure}[t]
\begin{minipage}{0.5\hsize}
\begin{center}
\includegraphics[height=35mm]{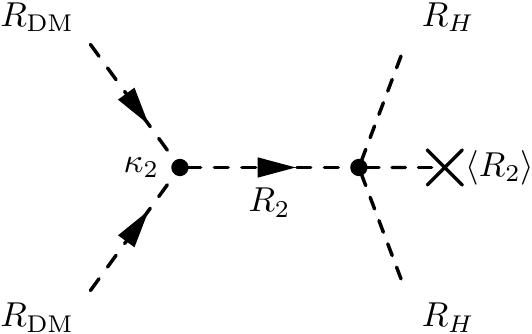}
\end{center}
\end{minipage}
\begin{minipage}{0.5\hsize}
\begin{center}
\includegraphics[height=35mm]{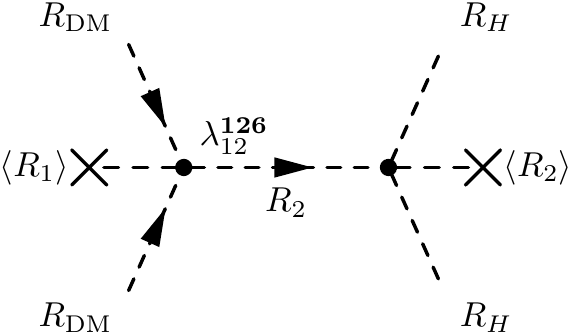}
\end{center}
\end{minipage}
\vspace{0.5cm}
\\
\begin{minipage}{0.5\hsize}
\begin{center}
\includegraphics[height=35mm]{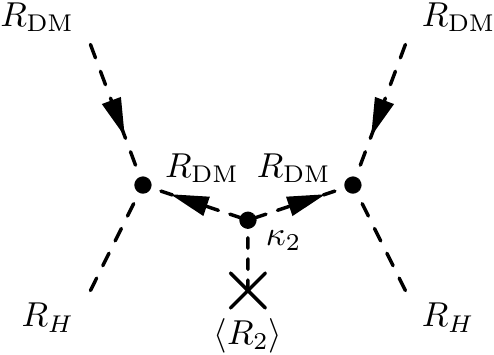}
\end{center}
\end{minipage}
\begin{minipage}{0.5\hsize}
\begin{center}
\includegraphics[height=35mm]{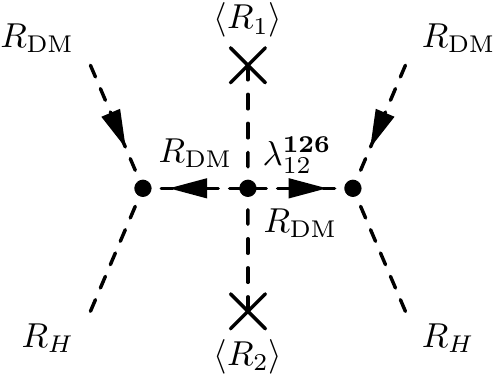}
\end{center}
\end{minipage}
\vspace{0.3cm}
\caption{{\it Diagrams that generate the mass splitting between real
 components of hypercharged scalar DM.}}
\label{fig:scatree}
\end{figure}

Such an operator is induced by the interactions with the coefficients
$\kappa_2$ and $\lambda_{12}^{\bf 126}$ in Eq.~\eqref{eq:lintsc}, since
it requires violation of the particle number associated with the DM field
$\phi$. The required $B-L$ breaking is realized by
the $R_2$ VEV. We find that the tree-level
diagrams in Fig.~\ref{fig:scatree} are relevant to the generation of
mass splitting. Here, $R_H = {\bf 10}$ contains the SM Higgs field. 
Since the $\kappa_2$ and $\lambda_{12}^{\bf 126}$ interactions are
symmetric with respect to the interchange of $R_{\text{DM}}$, the
component in $R_2$ which propagates in the upper two diagrams should be an
SU(2)$_L$ triplet. On the other hand, the component appearing in the
inner lines in the lower two diagrams can be either an SU(2)$_L$
singlet or triplet. The masses of these fields are dependent on the
intermediate gauge groups; if $G_{\text{int}} = \text{SU}(4)_C\otimes
\text{SU}(2)_L\otimes \text{SU}(2)_R$ or
$\text{SU}(3)_C\otimes \text{SU}(2)_L\otimes \text{SU}(2)_R \otimes
\text{U}(1)_{B-L}$, then these masses are ${\cal O}(M_{\text{GUT}})$,
while for $G_{\text{int}} =\text{SU}(3)_C\otimes \text{SU}(2)_L\otimes
\text{SU}(2)_R \otimes \text{U}(1)_{B-L}\otimes D$, they are ${\cal
O}(M_{\text{int}})$ because of the left-right parity $D$.

Let us first consider the former cases. In these cases, the coefficient of
the dimension-four operator $\phi^2 H^{*2}$ is ${\cal
O}(M_{\text{int}}/M_{\text{GUT}})$, as the dimensionful couplings in
the Lagrangian are expected to be ${\cal O}(M_{\text{GUT}})$. We note
here that there is no requirement for the cancellation between
$\kappa_2$ and $\lambda_{12}^{\bf 126}\langle R_1\rangle$ to realize the
desired mass spectrum since these couplings do not contribute to the
mass splitting as mentioned in footnote~\ref{ftnt:masssp}, in contrast
to the case we discuss below. Thus, this operator induces a mass
splitting of 
\begin{equation}
\Delta m \sim \frac{M_{\rm int}v^2}{m_{\text{DM}}M_{\rm GUT}} ~.
\end{equation}
The condition $\Delta m\gtrsim 100~\text{keV}$ then becomes
\begin{equation}
\frac{M_{\rm int}}{M_{\rm GUT}}\gtrsim 3 \times 10^{-6} \times
 \left(\frac{m_{\text{DM}}}{1~{\rm TeV}}\right) ~.
\end{equation} 
From Table~\ref{table:scalar_result1}, we find that the model \DM{SB}{422}
clearly satisfies this bound, while the intermediate scales in
\DM{SB}{3221} and \DM{SC}{3221} lie slightly below this
constraint. However, since this bound is just a rough estimation and the
intermediate scales given in Table~\ref{table:scalar_result1} are
obtained with the one-loop RGEs, it is possible that the DM candidates
in these models are just not yet reached by the current direct detection
experiments. If so, these models can be probed in the near future.

For \DM{SB}{3221D}, on the other hand,
the mass spectrum is altered because of the
presence of the left-right parity. In this case, the charge of the DM
candidate under $G_{\text{int}}$ is $({\bf 1}, {\bf 2}, {\bf 1}, -1)$,
and the left-right symmetry requires the $({\bf 1}, {\bf 1}, {\bf 2},
+1)$ to be also light. To that end, the fine-tuning between the
$\kappa_2$ and $\lambda_{12}^{\bf 126}$ terms in Eq.~\ref{eq:lintsc} is
required such that $\kappa_2 + \lambda_{12}^{\bf 126} \langle R_1
\rangle \simeq M_{\text{int}}$; otherwise, these terms give a mass of
${\cal O}(\sqrt{M_{\text{int}}M_{\text{GUT}}})$ to the $({\bf 1}, {\bf
1}, {\bf 2}, +1)$ component, which is much higher than the intermediate
scale. This fine-tuning also guarantees the absence of non-perturbative 
couplings at low energies; without this fine-tuning, the exchange
of intermediate-scale particles with the $\kappa_2$ and
$\lambda_{12}^{\bf 126} \langle R_1 \rangle$ vertices induces extremely
large effective couplings, which destroy the perturbativity of the
low-energy theory.  

In the presence of the fine-tuning, the diagrams in
Fig.~\ref{fig:scatree} with the virtual states having a mass of $M_{\rm
int}$ induce the effective operator $\phi^2 H^{*2}$ with a coefficient
of ${\cal O}(1)$. Thus, the resultant mass splitting is well above
100~keV and the model \DM{SB}{3221D} easily evades the constraints from
the direct detection experiments. 

To summarize, \DM{SB}{422} and \DM{SB}{3221D} are safe from the direct
detection bound. \DM{SB}{3221} and \DM{SC}{3221} lie just around the
margin of the bound, and they might be detected or completely excluded
in future direct detection experiments.


\subsection{Constraints and prospects for the scalar DM candidates}
\label{sec:scalarpheno}

The above discussions have revealed that the only possible scalar DM
candidates we could obtain with sufficiently high $M_{\text{GUT}}$
are \model{S}{1}{0} and \model{S}{2}{1/2}, as shown in
Table~\ref{table:scalar_result1}. Before concluding this section, we
briefly review the current constraints on these DM
candidates, and give prospects for probing them in future experiments.

First, we consider \model{S}{1}{0}. This DM candidate has been widely
discussed so far since it is one of the simplest extensions of the SM to
include a DM candidate \cite{Burgess:2000yq}. As we have seen in
Sec.~\ref{sec:scalarDMmass}, the thermal relic abundance of \model{S}{1}{0} is
determined once we fix the DM mass $m_{\text{DM}}$ and its quartic
coupling to the SM Higgs field, $\lambda_{H\phi}$. Therefore, by
requiring its thermal relic abundance to be equal to the observed DM
density $\Omega_{\text{DM}}h^2 = 0.12$, we can express $\lambda_{H\phi}$
as a function of the DM mass $m_{\text{DM}}$. Since this is the only
coupling that connects the DM to the SM sector, various physical
quantities relevant to the DM detection, such as the DM-nucleon
scattering cross section, are also determined in terms of the DM
mass. 

The present constraints on this DM are summarized in
Refs.~\cite{Cline:2013gha, Abe:2014gua}. According to those results, currently,  DM
direct detection experiments give a stringent limit on the DM mass; the
\model{S}{1}{0} DM with a mass of $m_{\text{DM}}\lesssim 53$~GeV and
$64~\text{GeV}\lesssim m_{\text{DM}} \lesssim 100$~GeV has been excluded
by the LUX experiment \cite{Akerib:2013tjd}. In addition, if
$m_{\text{DM}} < m_h/2$ with $m_h\simeq 125$~GeV the Higgs mass, then
the constraint on the invisible decay width of the Higgs boson also
restricts the DM. It turns out that the current upper bound on the
invisible decay width $\text{BR}(h\to \text{invisible})< 0.19$
\cite{Belanger:2013xza} leads to the limit on the DM mass of
$m_{\text{DM}} \gtrsim 53~\text{GeV}$. The DM direct
detection experiments with ton-scale detectors, such as XENON1T, will
probe most of the DM mass region, and thus this DM model can be tested
in the near future. 

DM described by model \model{S}{2}{1/2} is called the Inert
Higgs Doublet DM \cite{Deshpande:1977rw}, whose current status is summarized in
Refs.~\cite{Arhrib:2013ela, Abe:2014gua}. As discussed in these papers, 
favored mass regions for the \model{S}{2}{1/2} DM that account for the
correct DM abundance can be divided into two parts:
$m_{\text{DM}}\lesssim 100$~GeV and $m_{\text{DM}}\gtrsim 500$~GeV. In
the former case, the DM particles annihilate efficiently through the
Higgs boson exchange process, especially where $m_{\text{DM}}\simeq
m_h/2$. When $100~\text{GeV}\lesssim m_{\text{DM}} \lesssim
500~\text{GeV}$, the DM annihilation cross section is too large because
the $W^+W^-$ channel is open. For $m_{\text{DM}}\gtrsim 500$~GeV, both
the Higgs boson and the gauge bosons contribute to the DM annihilation
so that the resultant relic abundance can agree to the observed DM
density. For the lower mass region, the direct detection experiments,
the measurements of the Higgs decay branching ratios, and the
electroweak precision measurements restrict the parameter space. Both
of the mass regions can be probed in future direct detection
experiments \cite{Abe:2015rja}. Indirect detection experiments are, on
the other hand, less promising; still, depending on the DM profile,
gamma-ray searches from the Galactic Center may provide a signature
of this DM candidate.

\section{Fermionic dark matter}
\label{sec:fermionDM}

Next, we consider the fermionic DM candidates. Again, we begin with
showing the favored mass region for these DM candidates in
Sec.~\ref{sec:fermionDMmass}. As already mentioned above, the singlet
fermion candidates, \model{F}{1}{0} and \model{\widehat{F}}{1}{0}, are
not good candidates for WIMP DM since their annihilation cross sections
are extremely suppressed (though they are good NETDM candidates). 
On the other hand, electroweakly charged DM
can yield the desired relic abundance via gauge interactions. We
discuss the $Y=0$ and $Y\neq 0$ cases in
Sec.~\ref{sec:fermionrealtriplet} and
Sec.~\ref{sec:fermionhypercharged}, respectively. We give some concrete
examples for each case and perform RGE analysis to determine the
intermediate/GUT scales of the models. Finally, in
Sec.~\ref{sec:fermionpheno}, we summarize the present limits on these
fermion DM models and discuss future prospects for probing these DM
candidates.  Non-thermal hypercharged DM is discussed in Sec.~\ref{sec:NetHyperchargedDM}.

\subsection{DM mass}
\label{sec:fermionDMmass}

Contrary to the case of the scalar DM, the thermal relic abundance of
the fermionic DM candidates is completely determined by gauge
interactions. Therefore, it is possible to make a robust prediction for
the DM mass favored by the present DM density. In the case of fermion
DM, not only the gauge boson channels but also the SM fermions and the
Higgs boson final states can contribute to $s$-wave annihilation. We
obtain a similar expression to Eq.~\eqref{eq:effannscalar} for the
effective annihilation cross section of \model{F}{n}{Y} as \cite{Cirelli:2005uq}  
\begin{equation}
 \sigma_{\text{ann}}v_{\text{rel}} \simeq 
\frac{g^4 (2n^4+17n^2-19)
+4Y^2g^{\prime 4}(41+8Y^2)+16 g^2g^{\prime 2}Y^2 (n^2-1)}
{128\pi c_n m_{\text{DM}}^2} ~,
\end{equation}
with $c_n = 2n$ ($4n$) for a Majorana (Dirac) fermion. In addition, the
Sommerfeld enhancement again affects the annihilation cross section
significantly. With this effect taken into account, the thermal relic
abundance of \model{F}{3}{0} is computed in Ref.~\cite{Hisano:2006nn}
and found to be consistent with the observed DM density if
$m_{\text{DM}} \simeq 2.7$~TeV as in the case of supersymmetric winos. 
As for \model{F}{2}{1/2} and
\model{\widehat{F}}{2}{1/2}, the favored mass value is $\simeq 1.1$~TeV
\cite{Cirelli:2005uq} as in the case of supersymmetric Higgsinos. 
As far as we know, there is no calculation for
the other fermionic DM candidates that includes the Sommerfeld
enhancement; without the effect, the thermal relic of \model{F}{3}{1},
\model{\widehat{F}}{3}{1}, \model{F}{4}{1/2}, and \model{F}{4}{3/2} is
consistent with the observed value if $m_{\text{DM}}\simeq 1.9$~TeV,
1.9~TeV, 2.4~TeV, and 2.6~TeV, respectively \cite{Farina:2013mla}.

\subsection{Real triplet DM}
\label{sec:fermionrealtriplet}

We begin our discussion of fermionic DM models with the $Y=0$ case. As
discussed earlier, these are less constrained by direct
detection experiments. According to Table~\ref{tab:dmcandidate}, such
candidates belong to $\text{SU(2)}_L$ triplets in a $\bf 45$, $\bf 54$ or
$\bf 210$ of SO(10). A summary of $\text{SU}(4)_C\otimes \text{SU}(2)_L
\otimes \text{SU}(2)_R$ quantum numbers of these DM multiplets are listed
in Table.~\ref{tab:realtriplet}. Note that the $B-L$ and $T_R^3$
charges for all of these DM candidates vanish, and therefore they are
regarded as real Majorana fermions. As in the scalar DM scenario, the DM
multiplet in the $\bf 54$ or $\bf 210$ is degenerate with other components
with respect to $G_\text{int}$, and we are required to break this
degeneracy to avoid unwanted long-lived colored/charged particles
\cite{Mambrini:2015vna}. In the fermionic case, however, a
renormalizable Yukawa term like $\overline{R}_\text{DM}R_\text{DM} {\bf
126}_H$ is forbidden by SO(10) symmetry and the choice of DM
representation \cite{Mambrini:2015vna}, and thus we are unable to use the
{\bf 126} Higgs to break the degeneracy. Therefore, we need to introduce
additional Higgs fields at the intermediate scale in these cases.  

\begin{table}[ht!]
  \begin{center}
  \caption{\it Real triplet DM candidates in various SO(10) representations. }
  \label{tab:realtriplet}
  \vspace{5pt}
	\begin{tabular}{c | c}
	\hline
	\hline
	 ${\rm SO(10)}$ representation&  ${\rm SU(4)}_C\otimes{\rm
	 SU(2)}_L\otimes{\rm SU(2)}_R$  \\ 
        \hline
	${\bf 45}$ & $({\bf 1},{\bf 3},{\bf 1})$ \\
	${\bf 54}$ & $({\bf 1},{\bf 3},{\bf 3})$ \\
	${\bf 210}$ & $({\bf 15},{\bf 3},{\bf 1})$\\
	\hline
	\hline
	\end{tabular}
  \end{center}
\end{table}

For simplicity, we restrict ourselves to the cases where the
intermediate scale VEVs develop in the SM singlet direction of
$R_1$ and/or $R_2 ={\bf 126}$. One of the SM singlet components of $R_1$
should have a VEV of ${\cal O}(M_{\text{GUT}})$ to break SO(10) into
$G_{\text{int}}$. The $R_2$ Higgs field acquires an ${\cal
O}(M_{\text{int}})$ VEV to break $G_{\text{int}}$, but it is not able to
give mass differences among the components in $R_{\text{DM}}$, as
mentioned above. Thus, we need to exploit an extra SM singlet component in
$R_1$  which remains light compared to the GUT scale,
to induce intermediate-scale mass terms for $R_{\text{DM}}$, which
are to be used to generate the required mass splitting. We
denote the VEVs of these two components of $R_1$ which break SO(10) and
$G_\text{int}$ by $v_\text{GUT}\sim M_\text{GUT}$ and $v_\text{int}\sim
M_\text{int}$, respectively. Then, the mass splitting in the DM
multiplet $R_{\text{DM}}$ can be realized as follows:  
\begin{align}
-\mathcal{L}_\text{DM} &= M \overline{R}_\text{DM}R_\text{DM} 
- R_1\overline{R}_\text{DM}R_\text{DM}  \nonumber\\
&\rightarrow 
\left(M-c_1 v_\text{GUT} - c_2 v_\text{int}\right) \overline{\chi}\chi ~,
\label{eq:mintsplitf}
\end{align}
where $\chi$ denotes the DM field and $M\sim M_\text{GUT}$ is a
universal mass. $c_1$ and $c_2$ are the Clebsch-Gordan coefficients that
vary for different $R_\text{DM}$ components. Thus, by fine-tuning $M$
such that $M-c_1 v_\text{GUT} - c_2 v_\text{int} \sim 1$~TeV,
we can set the DM triplet to be at TeV scale while leaving other
contents in $R_\text{DM}$ either around $M_{\text{int}}$ or
$M_{\text{GUT}}$. 
We summarize in Table~\ref{tab:intR1} the multiplets in $R_1$ that may
develop a VEV of ${\cal O}(M_\text{int})$ for different
$G_\text{int}$. The multiplets are labeled by the quantum numbers of
$G_\text{int}$. It turns out that there is no extra SM singlet component
in {\bf 54}, which is indicated by a hyphen in the table. 
As a consequence, there is no way to fine-tune the mass of the ({\bf 1},
{\bf 3}, {\bf 3}) DM candidate originating from the {\bf 54} and we drop it from
further discussion. Here, we note
that the cases of $G_\text{int}=\text{SU(3)}_C\otimes \text{SU}(2)_L
\otimes \text{SU}(2)_R \otimes \text{U}(1)_{B-L}$ and
$\text{SU(3)}_C\otimes \text{SU}(2)_L \otimes \text{SU}(2)_R
\otimes \text{U}(1)_{B-L} \otimes D$ are disfavored before further
analysis: the addition of a real triplet DM lowers the unification scale to 
unacceptable values and 
in these cases there is neither any new-physics contribution
to the SU(3)$_C$ gauge coupling beta function nor any new positive
contribution to the SU(2)$_L$ beta function above $M_{\rm int}$.
Therefore, the
resultant $M_\text{GUT}$ is always smaller than the unification scale
of the SU(3)$_C$ and SU(2)$_L$ gauge couplings in the SM plus a real
triplet DM, which is below $10^{15}~{\rm GeV}$ and thus too low to evade
the proton decay constraint.\footnote{Note that scalar doublet DM is allowed under these intermediate symmetries as shown in Table~\ref{table:scalar_result1}, since its contribution to the beta functions is much smaller than that from a fermionic real triplet,
thus allowing for a higher unification scale.}  For this reason, we do not consider these
cases in Table~\ref{tab:intR1}. 

\begin{table}[ht!]
 \begin{center}
\caption{\it Possible components in $R_1$ that can develop a VEV of
  ${\cal O}(M_{\text{int}})$.}
\label{tab:intR1}
\vspace{5pt}
\begin{tabular}{llc}
\hline
\hline
$G_{\text{int}}$ & $R_1$ & Intermediate scale multiplets \\
\hline
{$\text{SU}(4)_C\otimes \text{SU}(2)_L \otimes \text{SU}(2)_R$}
& {\bf 210}&
$({\bf 15},{\bf 1},{\bf 1})$ \\ 
& & $({\bf 15},{\bf 1},{\bf 3})$\\
\rowcolor{LightGray}
$\text{SU}(4)_C\otimes \text{SU}(2)_L \otimes \text{SU}(2)_R\otimes {D}$
& {\bf  54}
& --
\\
$\text{SU}(4)_C\otimes \text{SU}(2)_L \otimes \text{U}(1)_R$ & {\bf 45}
&
{$({\bf 15},{\bf 1}, 0)$}
\\
\hline
\hline
\end{tabular}
 \end{center}
\end{table}

\begin{table}[ht!]
\centering
\caption{\it The one-loop results for $M_\text{GUT}$, $M_\text{int}$,
$\alpha_\text{GUT}$, and proton lifetimes for real triplet fermionic DM
 models. Here we set the DM mass to be 1~TeV. The mass scales and proton
 decay lifetime are in unit of GeV and years, respectively. In the blue
 shaded model, gauge coupling unification is achieved with a sufficiently
 high GUT scale. } 
\label{table:ftriplety0}
\vspace{5pt}
  \begin{tabular}{lccccc}
    \hline
    \hline
$R_{\rm DM}$ & Additional Higgs & $\mathrm{log}_{10} M_{\rm int} $ & 
$\mathrm{log}_{10} M_{\rm GUT} $ & $\alpha_{\rm GUT}$ &  $\log_{10}\tau_p(p\rightarrow e^+ \pi^0)$
 \\
 & in $R_1$ &&&& \\
 \hline 
\hline
 \multicolumn{6}{c}{$G_\text{int}={\rm SU(4)}_C\otimes{\rm SU(2)}_L\otimes{\rm
SU(2)}_R$}\\
\hline
$({\bf 1},{\bf 3},{\bf 1})$ & -- & $15.50$ & $13.69$ & $0.0263$ & --\\ 
\hline 
$({\bf 1},{\bf 3},{\bf 1})$ & $({\bf15},{\bf 1},{\bf 3})$ & -- & -- & -- & --\\ 
\hline 
$({\bf 1},{\bf 3},{\bf 1})$ & $({\bf15},{\bf 1},{\bf 1})$ & $15.65$ & $13.47$ & $0.0263$ & --\\ 
\hline 
 \rowcolor{LightBlue} 
$({\bf 1},{\bf 3},{\bf 1})$ & $({\bf15},{\bf 1},{\bf 1})$ & $ 6.54$ & $17.17$ & $0.0252$ & $39.8\pm 1.2$\\ 
 \rowcolor{LightBlue}
& $({\bf15},{\bf 1},{\bf 3})$ & & & & \\ 
\hline 
$({\bf15},{\bf 3},{\bf 1})$ & $({\bf15},{\bf 1},{\bf 1})$ & $14.44$ & $14.10$ & $0.0246$ & --\\ 
\hline 
$({\bf15},{\bf 3},{\bf 1})$ & $({\bf15},{\bf 1},{\bf 1})$ & $14.52$ & $14.11$ & $0.0243$ & --\\ 
& $({\bf15},{\bf 1},{\bf 3})$ & & & & \\ 
\bhline{1.5pt}
 \multicolumn{6}{c}{$G_\text{int}={\rm SU(4)}_C\otimes{\rm SU(2)}_L\otimes{\rm
SU(2)}_R\otimes D$}\\
\hline
$({\bf 1},{\bf 3},{\bf 1})$ & -- & $14.78$ & $14.04$ & $0.0250$ & --\\ 
\bhline{1.5pt}
 \multicolumn{6}{c}{$G_\text{int}={\rm SU(4)}_C\otimes{\rm SU(2)}_L\otimes{\rm
U(1)}_R$}\\
\hline
$({\bf15},{\bf 3},0)$ & $({\bf15},{\bf 1},{\bf 0})$ & $14.55$ & $14.21$ & $0.0246$ & --\\  

 \hline
 \hline
  \end{tabular}
\end{table}

We now perform the RG analysis to look for promising models with
additional intermediate Higgs multiplets given in Table~\ref{tab:intR1}.
The one-loop results for $M_\text{GUT}$, $M_\text{int}$,
$\alpha_\text{GUT}$, and proton lifetimes for different combination of
$R_\text{DM}$ and the Higgs fields are listed in
Table~\ref{table:ftriplety0}.\footnote{We again restrict our attention
to one-loop RGEs to avoid any model dependence due to the Yukawa
coupling with the additional Higgs in $R_1$.} Here, we set the DM mass
to be 1~TeV. The second column lists the extra Higgs fields in $R_1$ at $M_\text{int}$ in addition
to $R_2$. We suppressed combinations of Higgs multiplets that cannot split the degeneracy of DM multiplet as in Eq.~\eqref{eq:mintsplitf}.  The mass scales and proton decay lifetime are in units of GeV
and years, respectively. We find that there is only one promising model
with $G_\text{int}={\rm SU(4)}_C\otimes{\rm SU(2)}_L\otimes{\rm
SU(2)}_R$, which is highlighted by blue shading in Table.~\ref{table:ftriplety0}. In this case, since the DM multiplet is a singlet under both
$\text{SU(4)}_C$ and $\text{SU(2)}_R$, the additional Higgs fields are
not necessary from the viewpoint of the mass splitting for the DM
multiplet; namely, there is no degeneracy problem for this model. Rather, they are required
so that the model achieves a good unification scale beyond proton decay
constraint. The model has, however, a quite low intermediate scale that
results in large neutrino masses through the type-I seesaw mechanism
since the Dirac mass terms for neutrinos are related to the up-type
Yukawa couplings in this setup. A simple way to evade this problem is to
introduce a complex $({\bf 15},{\bf 2},{\bf 2})_C$ Higgs field in a {\bf 126} to
modify the relation, as discussed in
Ref.~\cite{Mambrini:2015vna}.\footnote{For the effects of a $({\bf
15},{\bf 2},{\bf 2})_C$ Higgs field on the Yukawa couplings, see
Refs.~\cite{Bajc:2005zf, Lazarides:1980nt}.} If a
$({\bf 15},{\bf 2},{\bf 2})_C$ Higgs is also present at the intermediate
scale, it turns out that gauge coupling unification is still realized,
with $\log_{10} M_{\text{int}}=9.28$, $\log_{10} M_{\text{GUT}}=16.38$,
$\alpha_{\text{GUT}} = 0.038$, and $\log_{10}\tau_p(p\rightarrow e^+
\pi^0) = 35.9$. Here again, the mass scales and proton decay lifetime
are expressed in units of GeV and years, respectively. Finally, we note
that the addition of $({\bf 15},{\bf 2},{\bf 2})_C$ will not resurrect
the failed models in Table~\ref{table:ftriplety0}.

\subsection{Hypercharged DM}
\label{sec:fermionhypercharged}

\begin{figure}[ht!]
\begin{center}
\subfigure[Scalar exchange]
 {\includegraphics[clip, height = 40mm]{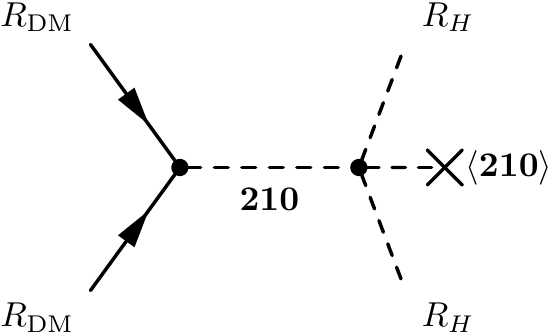}
 \label{fig:fertree1}}
\subfigure[Fermion exchange]
 {\includegraphics[clip, height=40mm]{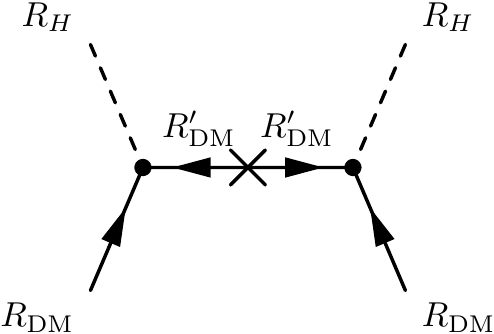}
 \label{fig:fertree2}}
\caption{{\it Diagrams that generate the mass splitting between the Weyl
 components of hypercharged Dirac DM through the exchange of an
 intermediate-scale (a) scalar (b) fermion.}}
\label{fig:msplitf}
\end{center}
\end{figure}

Hypercharged DM is a natural step forward after considering real triplet
DM. In this section, we still restrict the Higgs content as in the
previous section. As we discussed in Sec.~\ref{sec:ydmbound},
hypercharged DM is strongly constrained by direct detection
experiments. To evade this constraint, we need to split the mass of the
Weyl components of the hypercharged Dirac DM by $\sim
100~\text{keV}$. There are two possible ways to generate an effective
operator in Eq.~\eqref{eq:effmass} through exchange of a field at the
intermediate scale at tree level, depending on whether it is a scalar or
a fermion. In the former case, the effective operator is induced by the
exchange of intermediate-scale Higgs fields, as illustrated in
Fig.~\ref{fig:fertree1}.  This requires the hypercharge of the virtual
Higgs field to be at least one and $M_\text{int}\lesssim 10^9~\text{GeV}$.
According to Table~\ref{tab:intR1}, the only candidate for such a Higgs
field belongs to $(\bf{15},\bf{1},\bf{3})$ in the $\bf 210$ when
$G_\text{int}=\text{SU}(4)_C\otimes \text{SU}(2)_L \otimes
\text{SU}(2)_R$. The DM candidate should then be in a
$(\bf{15},\bf{2},\bf{2})$ or
$(\overline{\bf{10}},\bf{2},\bf{2})\oplus({\bf{10}},\bf{2},\bf{2})$
representation of $\text{SU}(4)_C\otimes \text{SU}(2)_L \otimes
\text{SU}(2)_R$. We performed a scan for models that contain above
content, and found that none of them gives appropriate $M_\text{int}$
and $M_\text{GUT}$. The latter possibility is to introduce another
fermionic real multiplet at the intermediate scale, so that the DM
candidate is a mixture of a hypercharged field and a Majorana
field. This mechanism is demonstrated in Fig.~\ref{fig:fertree2}, where
$R_\text{DM}$ is the main component of the DM candidate which is
hypercharged and has a mass term of TeV scale; $R^\prime_\text{DM}$ is
the Majorana field at the intermediate scale. The cross mark in
Fig.~\ref{fig:fertree2} represents the chiral flipping in the propagator of
the Majorana field $R^\prime_\text{DM}$. $R_\text{DM}$ and
$R^\prime_\text{DM}$ couple to the SM Higgs field through terms like  
\begin{equation}
\mathcal{L}_\text{mix}\propto\overline{R}_\text{DM}R^\prime_\text{DM}R_H+
\text{h.c.}
\label{majoranasplitf}  
\end{equation}
Since $R^\prime_\text{DM}$ is a Majorana field, it can only belong to
either a singlet or a real triplet among the possible candidates in
Table~\ref{tab:dmcandidate}. As a result, DM can only belong to a doublet
(\model{F}{2}{1/2} or \model{\widehat{F}}{2}{1/2}) or a quartet
(\model{F}{4}{1/2}), with hypercharge $1/2$. This requires
$M_\text{int}\lesssim 10^9~\text{GeV}$ according to the discussion in
Sec.~\ref{sec:ydmbound}. Note that the $Y\geq 1$ DM candidates,
\model{F}{3}{1}, \model{\widehat{F}}{3}{1}, and \model{F}{4}{3/2},
require at least $2 Y$ additional fermions at the intermediate scale to
generate the effective operator in Eq.~\eqref{eq:effmass}. To minimize
our model content, we do not consider these possibilities in the
following discussion. 

Taking the above discussion into account, we list the possible SO(10)
representations for $R_\text{DM}$ in the upper part of
Table~\ref{tab:doubletsinglet}; the singlet and real triplet candidates
for $R^\prime_\text{DM}$ are listed in the lower part of
Table~\ref{tab:doubletsinglet} and Table~\ref{tab:realtriplet},
respectively. The quantum numbers of the DM candidates with respect to
the intermediate gauge groups we consider can be inferred from the ${\rm
SU(4)}_C \otimes{\rm SU(2)}_L\otimes{\rm SU(2)}_R$ and $B-L$ quantum
numbers listed in the table.  

\begin{table}[ht!]
  \begin{center}
  \caption{\it The upper half of the table shows the fermionic $Y=1/2$
   candidates for $R_\text{DM}$ in various SO(10) representations; the
   lower half of the table shows the fermionic singlet candidates for
   $R^\prime_\text{DM}$.} 
  \label{tab:doubletsinglet}
  \vspace{5pt}
	\begin{tabular}{l  l l}
	\hline
	\hline
	 ${\rm SO(10)}$ representation &  ${\rm SU(4)}_C\otimes{\rm
	 SU(2)}_L \otimes{\rm SU(2)}_R$ & $B-L$ \\
        \hline
        ${\bf 10}$, ${\bf 120}$, ${\bf 210'}$ & $({\bf 1},{\bf 2},{\bf 2})$ & 0\\
	${\bf 120}$, ${\bf 126}$ & $({\bf 15},{\bf 2},{\bf 2})$ & 0\\
	${\bf 210}$ & $({\bf 10},{\bf 2},{\bf 2})\oplus(\overline{\bf 10},{\bf 2},{\bf 2})$ & $\pm 2$\\
	${\bf 210'}$ & $({\bf 1},{\bf 4},{\bf 4})$ & 0 \\
	\hline
	${\bf 54}$, ${\bf 210}$ & $({\bf 1},{\bf 1},{\bf 1})$ & 0\\
	${\bf 45}$             & $({\bf 1},{\bf 1},{\bf 3})$ & 0\\
	${\bf 45}$, ${\bf 210}$ & $({\bf 15},{\bf 1},{\bf 1})$ & 0\\
	${\bf 210}$            & $({\bf 15},{\bf 1},{\bf 3})$ & 0\\
	${\bf 126}$            & $({\bf 10},{\bf 1},{\bf 3})$ & 2\\
	\hline
	\hline
	\end{tabular}
  \end{center}
\end{table}


Now, we perform a one-loop calculation of $M_\text{int}$, $M_\text{GUT}$
and the proton decay lifetime for various combination of $R_\text{DM}$,
$R^\prime_\text{DM}$ and intermediate scale Higgs fields. Then, we pick
up the models that are not yet ruled out by proton decay experiments,
and at the same time have a relatively low intermediate scale
$M_\text{int}\lesssim 10^9$. We also require that the models have
appropriate particle and Higgs content, so that the DM acquires the
right mass through Eq.~\eqref{eq:mintsplitf} and
Eq.~\eqref{majoranasplitf}. It turns out that the viable models are
limited to $G_\text{int}={\rm SU(4)}_C\otimes{\rm SU(2)}_L\otimes{\rm
SU(2)}_R$ or  ${\rm SU(4)}_C\otimes{\rm SU(2)}_L \otimes{\rm
U(1)}_R$. These models are listed in Table~\ref{tab:hyperchargedfdm}
and no quartic models (\model{F}{4}{1/2}) were found.
The model \DM{FA}{422} is incompatible with small neutrino masses,
since the Yukawa coupling for the $\bf 16$ of this model is unified at
$M_\text{GUT}$. For models \DM{FA}{421} and \DM{FB}{422} , on the other
hand, we can avoid the neutrino mass problem by fine-tuning the Yukawa
couplings with additional Higgs fields at the intermediate scale, as
discussed in Sec.~\ref{sec:fermionrealtriplet}. Among them, the model
\DM{FA}{421} has a phenomenologically interesting consequence. Since
$M_{\text{int}} \simeq 3$~TeV, this model predicts a new massive neutral
gauge boson, $Z^\prime$, and vector leptoquarks 
whose masses are around a few TeV. These particles can 
be probed in future LHC experiments; for instance, dilepton resonance
searches \cite{Khachatryan:2014fba} are powerful probes for such a
$Z^\prime$. The leptoquarks are pair produced at the LHC, and their
signature is observed in dijet plus dilepton channels
\cite{Aad:2015caa}. Since they are produced via the strong interaction,
their production cross section is quite large. Thanks to the distinct
final states and large production cross section, the LHC experiments can
probe TeV-scale leptoquarks at the next stage of the LHC running.

\begin{table}[ht!]
\centering
\caption{\it \label{tab:hyperchargedfdm}
Possible hypercharged fermionic DM models that is not yet excluded by
 current proton decay experiments. The quantum numbers are labeled in
 the same order as $G_\text{int}$. The subscripts D and W refer
 to Dirac and Weyl respectively.  The numerical results are calculated
 for DM mass of 1~TeV. The mass scales and proton decay lifetime are in
 unit of GeV and years, respectively. }
  \vspace{5pt}
\scalebox{0.92}{
  \begin{tabular}{l l  l  l  l  l  l l }
    \hline
    \hline
 Model & $R_{\rm DM}$ & $R^\prime_{\rm DM}$ & Higgs & $\mathrm{log}_{10} M_{\rm int} $ & 
$\mathrm{log}_{10} M_{\rm GUT} $ & $\alpha_{\rm GUT}$ &
   $\log_{10}\tau_p $
 \\
\hline
 \hline
 \multicolumn{8}{c}{$G_\text{int}={\rm SU(4)}_C \otimes{\rm SU(2)}_L\otimes{\rm U(1)}_R$}
 \\
 \hline 
 \rowcolor{LightBlue}
 \DM{FA}{421} & $({\bf 1},{\bf 2},1/2)_D$ 
 & $({\bf 15},{\bf 1},0)_W$ 
 & $({\bf 15},{\bf 1},0)_R$ 
 & $3.48$  & $17.54$ & $0.0320$ 
 & $40.9\pm 1.2$ \\
 \rowcolor{LightBlue}
 &&& $({\bf 15},{\bf 2},{1}/{2})_C$&&&& \\
 \bhline{1.5pt}
  \multicolumn{8}{c}{$G_\text{int}={\rm SU(4)}_C\otimes{\rm
   SU(2)}_L\otimes{\rm SU(2)}_R$} 
 \\
 \hline 
 \rowcolor{LightBlue}
 \DM{FA}{422}& $({\bf 1},{\bf 2},{\bf 2})_W$ 
 & $({\bf 1},{\bf 3},{\bf 1})_W$ 
 & $({\bf 15},{\bf 1},{\bf 1})_R$ 
 & $9.00$  & $15.68$ & $0.0258$ 
 & $34.0\pm 1.2$ \\
 \rowcolor{LightBlue}
&&& $({\bf 15},{\bf 1},{\bf 3})_R$&&&& \\
 \hline 
\rowcolor{LightBlue}
 \DM{FB}{422}& $({\bf 1},{\bf 2},{\bf 2})_W$ 
 & $({\bf 1},{\bf 3},{\bf 1})_W$ 
 &  $({\bf 15},{\bf 1},{\bf 1})_R$
 & $5.84$  & $17.01$ & $0.0587$ 
 & $38.0\pm 1.2$ \\
  \rowcolor{LightBlue}
&&&$({\bf 15},{\bf 2},{\bf 2})_C$ &&&& \\
 \rowcolor{LightBlue}
&&& $({\bf 15},{\bf 1},{\bf 3})_R$ &&&& \\
 \hline
 \hline
  \end{tabular}
}
\end{table}

\begin{figure}[t]
\begin{center}
\includegraphics[height=80mm]{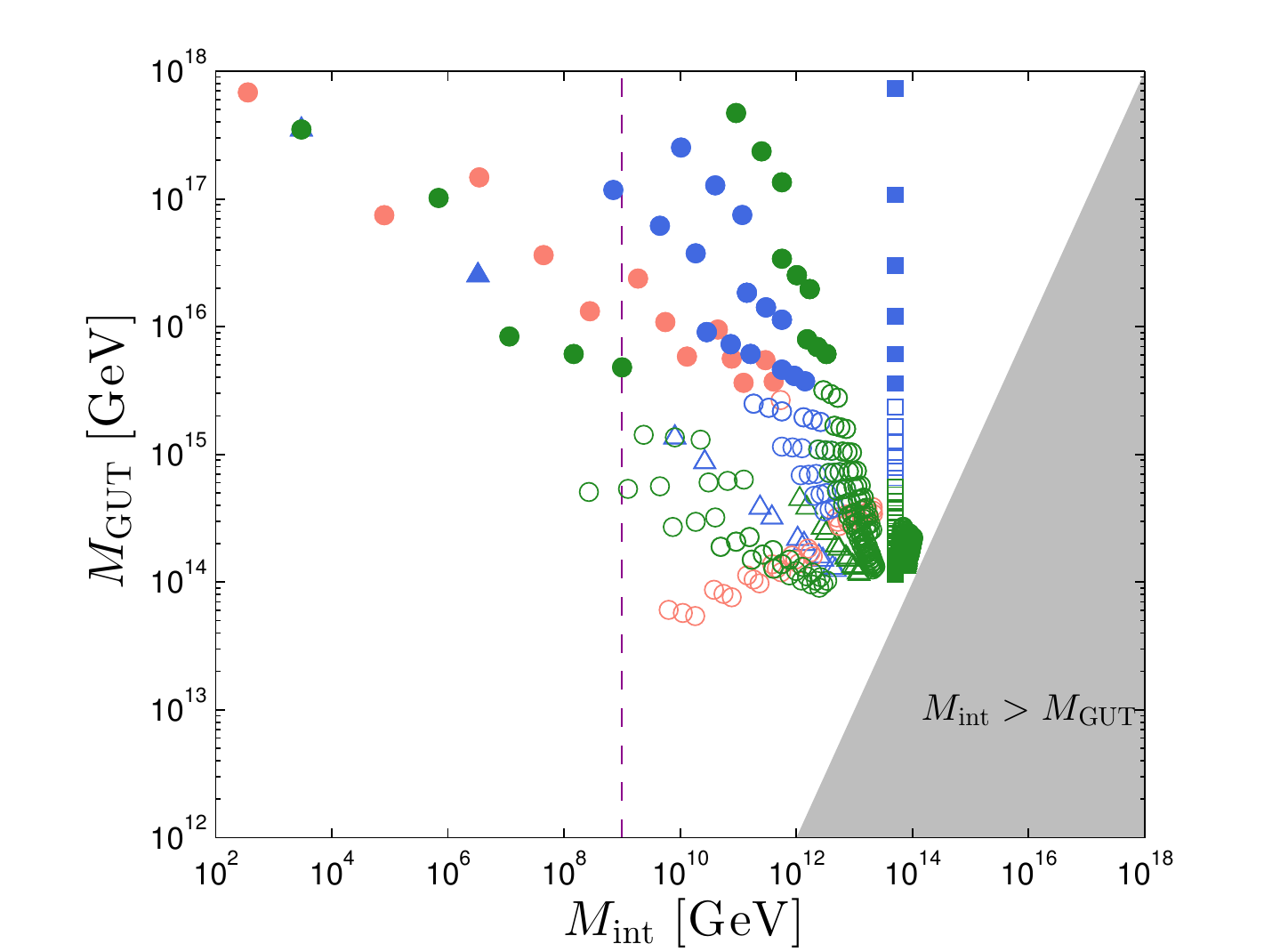}
\end{center}
\vspace{0.3cm}
\caption{{\it A scatter plot of general fermionic DM models. Here, the
 DM mass is set to be 1~TeV. The real triplet DM, doublet-singlet mixing
 DM and doublet-triplet mixing DM cases are colored in red, blue and
 green, respectively. The triangle, circle and square marker corresponds
 to $G_\text{int}={\rm SU(4)}_C \otimes{\rm SU(2)}_L\otimes{\rm
 U(1)}_R$, ${\rm SU(4)}_C\otimes{\rm SU(2)}_L\otimes{\rm SU(2)}_R$ and
 ${\rm SU(4)}_C\otimes{\rm SU(2)}_L\otimes{\rm SU(2)}_R\otimes D$,
 respectively. The vertical dashed line at $10^9~\text{GeV}$ indicates the direct detection constraint for $Y=1/2$ dark matter. The gray shaded area is disfavored for
 $M_\text{int}>M_\text{GUT}$. Only the filled symbols are consistent with a sufficiently 
 long proton lifetime.}}  
\label{fig:scatterplot}
\end{figure}

To conclude this section, we perform a scan for more general models
where the additional intermediate scale Higgs fields are not restricted
to the ones in $R_1$. Instead, they can be any combination of
$G_\text{int}$ representations that contain SM singlets. The Higgs fields can be taken to be either real or
complex. Moreover, we also consider the possible addition of a $({\bf 15},{\bf 2},{\bf 2})_C$
Higgs at the intermediate scale, which can be used to evade the problem of large
neutrino masses. The result of the scan is demonstrated in a scatter plot in
Fig.~\ref{fig:scatterplot}. The DM mass is again fixed to be 1~TeV. The
real triplet DM, $R_{\rm DM}$--$R^\prime_{\rm DM}$ doublet-singlet mixing DM and doublet-triplet mixing DM
cases are colored in red, blue and green, respectively. The triangle,
circle and square marker corresponds to $G_\text{int}={\rm SU(4)}_C
\otimes{\rm SU(2)}_L\otimes{\rm U(1)}_R$, ${\rm SU(4)}_C\otimes{\rm
SU(2)}_L\otimes{\rm SU(2)}_R$ and ${\rm SU(4)}_C\otimes{\rm
SU(2)}_L\otimes{\rm SU(2)}_R\otimes D$, respectively. The
$M_\text{int}>M_\text{GUT}$ region is theoretically disfavored, and is
indicated by the gray shaded area. In this plot, we do not consider the
realizability of the mass hierarchy for the DM multiplet, and thus the
number of good models should be smaller than that shown in the plot. 
All of the ${\rm SU(4)}_C\otimes{\rm SU(2)}_L\otimes{\rm SU(2)}_R\otimes
D$ cases with doublet DM predict the same $M_{\text{int}}$, since the
addition of extra fields at the intermediate scale does not change
$M_{\text{int}}$ in the presence of the left-right symmetry
\cite{Mambrini:2015vna}. As
can be seen from the figure, model points are concentrated in the high
intermediate scale and low GUT scale region. After we apply the
constraint of proton decay lifetime (shown by the filled symbols) 
as well as the condition
$M_\text{int}\lesssim 10^9~\rm{GeV}$ for the doublet DM cases, the
viable candidates turn out to be fairly limited.

  
\subsection{Constraints and prospects for the fermion DM candidates}
\label{sec:fermionpheno}

Finally, we review the present experimental constraints and future
prospects of the fermionic DM candidates discussed in this section. Let
us begin with the real triplet DM case. At the LHC, this DM candidate can be
probed by searching for disappearing tracks caused by the charged
component of the triplet DM, which has a decay length of ${\cal
O}(1)$~cm. Such a small decay length is due to the small mass difference
between the neutral and charged components; it is as small as a
hundred MeV, since it is induced at loop level\footnote{Currently, the
mass difference is computed at two-loop level \cite{Ibe:2012sx}: for a
3-TeV triplet DM, the mass difference is about 165~MeV.} after electroweak
symmetry breaking. Based on this search strategy, the ATLAS experiment
has searched for triplet DM and has given a lower bound
on the DM mass of $m_{\text{DM}}>270$~GeV \cite{Aad:2013yna}. For future
prospects on collider searches of triplet DM, see
Ref.~\cite{Shingo:2013aja}. Indirect searches of triplet DM are also
promising since this DM has a large annihilation cross section, as
already seen in Sec.~\ref{sec:fermionDMmass}. Indeed, an excess of
cosmic-ray antiprotons observed by the AMS-02 experiment \cite{ams02}
might be the first signature of triplet DM \cite{Ibe:2015tma}. 
On the other hand, this DM is currently
being constrained by the searches for gamma-ray line spectrum coming
from the Galactic Center. As discussed in Ref.~\cite{Cohen:2013ama},
the results from the H.E.S.S. Collaboration
\cite{Abramowski:2013ax} may give a strong limit on triplet DM, though
the consequences depend on the DM density profile used in
the analysis. More robust constraints are obtained by means of the
observation of gamma rays from dwarf spheroidal galaxies given by the
Fermi-LAT collaboration \cite{Ackermann:2013yva}; according to that
result, the mass of triplet DM is limited to $320~\text{GeV}\leq
m_{\text{DM}}\leq 2.25~\text{TeV}$ and $2.43~\text{TeV}\leq
m_{\text{DM}}$ at 95\% confidence level \cite{Bhattacherjee:2014dya}. In the
future, gamma-ray search 
experiments can probe a wider range of masses region for triplet
DM. Direct detection experiments are also able to catch the signature of
triplet DM. The scattering of triplet DM with a nucleon is
induced by the exchange of the electroweak gauge bosons at loop level
\cite{Hisano:2004pv}, and its scattering cross section is evaluated at
the next-to-leading order in Ref.~\cite{Hisano:2015rsa}: for instance,
$\sigma_{\text{SI}}\simeq 2\times 10^{-47}~\text{cm}^2$ for a 3~TeV real
triplet DM, which is well above the neutrino background \cite{Billard:2013qya}. For relevant works, see
also Ref.~\cite{Hill:2011be}. As a consequence, the triplet DM scenario
can be tested in various future experiments, and therefore is a quite
interesting possibility among the SO(10) DM candidates. 

Next, we consider the doublet DM case. Contrary to triplet DM,
doublet DM is rather hard to probe in experiments. In this case, the
mass difference between the neutral and charged components is found to be
as large as several hundreds of MeV, which makes it difficult to search
for doublet DM signal events at the LHC by using the disappearing
track method. The most promising way to probe doublet DM is the 
direct pair production at the ILC, which also enables us to study its
properties precisely \cite{Baer:2014yta}. The indirect DM searches are also
less promising due to a relatively small annihilation cross section of
doublet DM. The direct detection of this DM is only possible when
the intermediate scale is as low as $10^5$~GeV \cite{Nagata:2014aoa}; in
this case, the DM-nucleon scattering occurs through the exchange of the
Higgs boson, which is induced by effective operators generated at the
intermediate scale.\footnote{The electroweak loop contribution to the
DM-nucleon scattering in this case turns out to be very small
\cite{Hisano:2015rsa}. } In addition, if there are additional CP phases
in the effective operators, the electric dipole moment of electron is
sensitive to the effects of doublet DM \cite{Nagata:2014wma,
Nagata:2014aoa}. For instance, the model \DM{FA}{421} and \DM{FB}{422} in
Table~\ref{tab:hyperchargedfdm} may be tested in these
experiments. After all, the prospects for probing doublet DM 
quite depend on the intermediate scale, and future experiments are able
to search for some of the doublet DM models discussed above.

\subsection{Non-thermal Hypercharged DM}
\label{sec:NetHyperchargedDM}

In our previous discussion of fermionic hypercharged DM, we have assumed
the DM to be in thermal equilibrium before freeze out so that the DM
mass is restricted to be $\mathcal{O}(1~\rm{TeV})$ by the observed
abundance. As a consequence, hypercharged DM is highly restricted by
direct detection experiments. One way around such a constraint is to
introduce a small mass splitting between the Weyl components of the DM
by mixing the DM with another Majorana multiplet at the intermediate
scale. On the other hand, as suggested in Ref.~\cite{Feldstein:2013uha},
non-thermally produced DM can be extremely heavy and thus avoid the
direct detection constraint even when there is no mass splitting between
the Weyl components.  As a result, a minimal hypercharged DM model is
possible in this scenario. This minimality motivates us to consider this
class of DM candidates, even though they are not WIMPs. 

First, let us derive the lower bound on the DM mass to evade the direct
detection constraints. A fermionic DM particle with non-zero hypercharge $Y$
scatters off a nucleus via the exchange of $Z$ boson, and its scattering
cross section is given by 
\begin{equation}
 \sigma =\frac{G_F^2 Y^2}{2\pi}[N-(1-4 \sin^2\theta_W)Z]^2 \left(
\frac{m_{\text{DM}} m_T}{m_{\text{DM}} +m_T}
\right)^2 ~,
\end{equation}
where $G_F$ is the Fermi constant, $\theta_W$ is the weak mixing angle,
$Z$ and $N$ are the numbers of protons and neutrons in the nucleus,
respectively, and $m_T$ denotes the mass of the target nucleus. The LUX limit
\cite{Akerib:2013tjd} then reads\footnote{For ${}^{131}$Xe target,
$Z=54$, $N=77$.}
\begin{equation}
m_\text{DM}\gtrsim \left(2Y\right)^2\times 6\times 10^7~\rm{GeV}~.
\label{eq:directdetectionbound}
\end{equation} 

Such a heavy DM candidate can lead to the correct relic abundance if its mass
is larger than the reheating temperature $T_R$ after inflation so that
it never reaches equilibrium with the thermal bath of SM particles. Then
by carefully choosing the DM mass, the reheating temperature, and the
maximum temperature after inflation, one obtains the desired relic
abundance. The reheating temperature was shown to be in a range of
\cite{Feldstein:2013uha} 
\begin{equation}
T_R\simeq 10^{(\text{7--9})} \left( \frac{m_\text{DM}}{3\times 10^{10}~\text{GeV}} \right)~.
\end{equation}

In this scenario, it is natural to assume the DM mass scale to be equal
to intermediate scales of the unification models that we
consider. Usually, $M_\text{int}$ is large enough to evade the direct
detection bound and yet not so large that gravitational production of DM
becomes dominant. Moreover, we do not need to worry about mass splitting
in the DM multiplet, as we did in the WIMP scenario. For such heavy
particles, the electroweak and strong corrections to the masses of
charged or colored particles in the multiplet are large enough to
prevent them from acquiring cosmological lifetimes. This allows us to
consider minimal hypercharged DM models, which contain only one DM
multiplet without any extra Higgs multiplets with respect to the minimal
SO(10) unification model.

The  $Y=1/2$ DM candidates we consider include those in the upper
part of Table~\ref{tab:doubletsinglet}, and the $Y=1$ candidates listed in
Table~\ref{tab:hyperchargedtriplets}. The last multiplet in the upper
part of Table~\ref{tab:doubletsinglet} also contains an $Y=3/2$
candidate of DM which we will also consider. Now that the DM
multiplet only contributes to the running above $M_\text{int}$, we will
also consider the cases of $G_\text{int}=\text{SU(3)}_C\otimes
\text{SU}(2)_L \otimes \text{SU}(2)_R \otimes \text{U}(1)_{B-L}$ and
$\text{SU(3)}_C\otimes \text{SU}(2)_L \otimes \text{SU}(2)_R
\otimes \text{U}(1)_{B-L} \otimes D$, in contrast to the WIMP
scenario. 

\begin{table}[ht!]
  \begin{center}
  \caption{\it Fermionic $Y=1$
   candidates for $R_\text{DM}$ in various SO(10) representations;} 
  \label{tab:hyperchargedtriplets}
  \vspace{5pt}
	\begin{tabular}{l  l l}
	\hline
	\hline
	 ${\rm SO(10)}$ representation &  ${\rm SU(4)}_C\otimes{\rm
	 SU(2)}_L \otimes{\rm SU(2)}_R$ & $B-L$ \\
        \hline
        ${\bf 54}$ & $({\bf 1},{\bf 3},{\bf 3})$ & 0\\
	${\bf 126}$ & $({\bf 10},{\bf 3},{\bf 1})$ & 2\\
	\hline
	\hline
	\end{tabular}
  \end{center}
\end{table}

After an exhaustive calculation for different choices of
$R_\text{DM}$ and $G_\text{int}$, we find several possible models listed
in Table~\ref{tab:NetHyperchargedModels} that survive the direct
detection and proton decay constraints. Viable minimal models only exist for doublet DM when $G_\text{int}={\rm SU(4)}_C \otimes{\rm SU(2)}_L\otimes{\rm SU(2)}_R$, $G_\text{int}={\rm SU(4)}_C\otimes{\rm
   SU(2)}_L\otimes{\rm SU(2)}_R\otimes D$ and $G_\text{int}={\rm SU(3)}_C\otimes{\rm
 SU(2)}_L\otimes{\rm SU(2)}_R\otimes {\rm U(1)}_{B-L}$.  The
 intermediate scale of each model is larger than $10^8$~GeV, large
 enough to evade the direct detection bound indicated by
 Eq.~\eqref{eq:directdetectionbound}. As can be seen, most of the
 models in Table~\ref{tab:NetHyperchargedModels} can be probed in
 future proton decay experiments, though it is hard to detect these DM
 candidates in direct detection experiments. 

%
\begin{table}[ht!]
\centering
\caption{\it \label{tab:NetHyperchargedModels}
Possible non-thermal hypercharged fermionic DM models that are not yet
 excluded by current direct detection and proton decay experiments. The
 quantum numbers are labeled in the same order as $G_\text{int}$. The
 numerical results are calculated for DM mass equal to $M_{\rm int}$. The mass scales
 and proton decay lifetimes are in  unit of GeV and years, respectively. }
  \vspace{5pt}
\scalebox{0.92}{
  \begin{tabular}{l  l  l  l  l l }
    \hline
    \hline
 Model & $R_{\rm DM}$  & $\mathrm{log}_{10} M_{\rm int} $ & 
$\mathrm{log}_{10} M_{\rm GUT} $ & $\alpha_{\rm GUT}$ &
   $\log_{10}\tau_p $ \\
\hline
\hline
 \multicolumn{6}{c}{$G_\text{int}={\rm SU(4)}_C \otimes{\rm SU(2)}_L\otimes{\rm SU(2)}_R$}\\
 \rowcolor{LightBlue}
 \DM{FNA}{422} & $({\bf 1},{\bf 2},{\bf 2})_W$ 
 & $12.10$  & $15.63$ & $0.0225$ 
 & $34.0\pm 1.2$ \\
  \rowcolor{LightBlue}
 \DM{FNB}{422} & $({\bf 15},{\bf 2},{\bf 2})_W$ 
 & $11.15$  & $16.77$ & $0.0387$ 
 & $37.9\pm 1.2$ \\
 \bhline{1.5pt}
  \multicolumn{6}{c}{$G_\text{int}={\rm SU(4)}_C\otimes{\rm
   SU(2)}_L\otimes{\rm SU(2)}_R\otimes D$} \\
 \rowcolor{LightBlue}
 \DM{FNA}{422D} & $({\bf 15},{\bf 2},{\bf 2})_W$ 
 & $13.71$  & $15.36$ & $0.0286$ 
 & $32.8\pm 1.2$ \\
  \rowcolor{LightBlue}
 \DM{FNB}{422D} & $({\bf 10},{\bf 2},{\bf 2})_D$ 
 & $13.71$  & $15.94$ & $0.0342$ 
 & $34.9\pm 1.2$ \\
 \bhline{1.5pt}
  \multicolumn{6}{c}{$G_\text{int}={\rm SU(3)}_C\otimes{\rm
   SU(2)}_L\otimes{\rm SU(2)}_R\otimes {\rm U(1)}_{B-L}$} \\
 \rowcolor{LightBlue} 
 \DM{FNA}{3221} & $({\bf 1},{\bf 2},{\bf 2},0)_W$ 
 & $10.34$  & $15.82$ & $0.0227$ 
 & $34.8\pm 1.2$ \\ 
 \hline
 \hline
  \end{tabular}
}
\end{table}



\section{Conclusion and discussion}
\label{sec:conclusion}

The success of the Standard Model is now well established.
Nevertheless, we know that the Standard Model is incomplete.
Neutrinos have masses, there is a non-zero baryon asymmetry in the Universe,
and dark matter makes up a sizable component of the total matter density.
Many extensions to the Standard Model have been studied to explain
these phenomena. But rarely can a single extension explain all three.
SO(10) grand unification is one such example.

In most models, SO(10) symmetry breaking involves an intermediate 
gauge group, whose unknown scale presumably lies between the weak scale
and the grand unified scale (defined by the renormalization scale
where the gauge couplings are equal).  Standard Model fermions are neatly
contained in a {\bf 16} of SO(10) which includes all of the known SM fermions
plus a right-handed neutrino per generation. As the  right-handed neutrino 
is a SM singlet, it easily picks up a mass of order the intermediate scale
during the second phase of symmetry breaking, and leads to a natural
realization of the see-saw mechanism \cite{Minkowski:1977sc} for the
generation of neutrino masses. If produced (thermally or non-thermally) after
inflation, these same right-handed neutrinos can decay out of
equilibrium and produce a lepton asymmetry which can be converted
through electroweak effects to a baryon asymmetry, a process known as
leptogenesis \cite{fyl}. Furthermore the existence of an intermediate
scale makes gauge coupling unification feasible in the absence of
supersymmetry. 

Here we studied in detail one of the often unheralded features of SO(10) grand unification.
Namely its ability to provide for a WIMP dark matter candidate in addition to 
the benefits described above. Since SO(10) includes an additional U(1) symmetry beyond
hypercharge, if the 
Higgs field that breaks this symmetry at the intermediate scale belongs to a ${\bf
126}$ dimensional 
representation, then a discrete $\mathbb{Z}_2$
symmetry is preserved at low energies. This discrete symmetry (equivalent to matter parity
$P_M=(-1)^{3(B-L)}$) naturally provides the stability
for dark matter candidates. We considered all possible intermediate gauge group
with broken SU(5).

Stable SO(10) scalar (fermion) DM candidates must be odd (even) under the $\mathbb{Z}_2$ symmetry.
Therefore fermions must originate in either a  {\bf 10}, {\bf 45}, {\bf 54},  {\bf 120},  {\bf 126}, {\bf 210}
or  {\bf 210$^\prime$} representation, while scalars are restricted to either a  {\bf 16} or  {\bf 144} of SO(10).
These multiplets must be split and we gave explicit examples of fine-tuning mechanisms
in order to retain a 1 TeV WIMP candidate which may be a 
SU(2)$_L$, singlet, doublet, triplet, or quartet with or without hypercharge. 
Fermions which are SU(2)$_L$ singlets with no hypercharge are not good WIMP candidates but are NETDM candidates
and these were considered elsewhere \cite{Mambrini:2015vna}. 
Our criteria for a viable dark matter model required:
gauge coupling unification at a sufficiently high scale to ensure proton stability compatible with
experiment; a unification scale greater than the intermediate scale; and elastic cross sections
compatible with direct detection experiments. The latter criterion often requires additional 
Higgs representations to split the degeneracy of the fermionic
intermediate scale representations if DM is hypercharged. 

Despite the potential very long list of candidates (when one combines the possible
different SO(10) representations and intermediate gauge groups),
we found only a handful of models which satisfied all constraints.
Among the scalar candidates, the $Y=0$ singlet and $Y=1/2$ doublet (often referred to as
an inert Higgs doublet \cite{Deshpande:1977rw})
are possible candidates for $\text{SU}(4)_C\otimes \text{SU}(2)_L\otimes \text{SU}(2)_R$
and $\text{SU}(3)_C\otimes \text{SU}(2)_L\otimes \text{SU}(2)_R \otimes
 \text{U}(1)_{B-L}$ (with or without a left-right symmetry) intermediate gauge groups.
 These originate from either the {\bf 16} or {\bf 144} of SO(10).
 The latter group (without the left-right symmetry) is also consistent with a state originating 
 from the {\bf 144} being a triplet under SU(2)$_R$. 
 To avoid immediate exclusion from direct detection experiments, a mass splitting
 of order 100~keV implies that the intermediate scale must be larger than about 3 $\times 10^{-6}$~$M_{\rm GUT}$ for a nominal 1~TeV hypercharged scalar DM particle.
 Some of these models imply proton lifetimes 
 short enough to be testable in on-going and future proton decay experiments. 

The fermion candidates were even more restrictive. Models with $Y=0$
must come from a SU(2)$_L$ triplet (singlets are not WIMPs). In this case
only one model was found using the $\text{SU}(4)_C\otimes \text{SU}(2)_L\otimes \text{SU}(2)_R$
intermediate gauge group and requiring additional Higgses (already present in $R_1$) at 
the intermediate scale. Models with $Y=1/2$ doublets were found for $\text{SU}(4)_C\otimes \text{SU}(2)_L\otimes \text{U}(1)_R$ with a singlet fermion required for mixing, and $\text{SU}(4)_C\otimes \text{SU}(2)_L\otimes \text{SU}(2)_R$ with a triplet fermion for mixing. In both cases, additional Higgses from $R_1$ are
required at the intermediate scale.  More possibilities can be found if the additional 
Higgs are taken outside $R_1$. 

SO(10) almost always involves rather large representations (at least when compared with 
minimal SU(5) for example). SO(10) itself can be broken by either a {\bf 45}, {\bf 54} or {\bf 210}
representation ($R_1$) which determines the intermediate scale gauge group and 
a {\bf 126} is needed to break down to the SM (and preserve the needed $\mathbb{Z}_2$ symmetry).
The SM Higgs originates from a {\bf 10} and matter fields reside in three copies of  {\bf 16}'s. 
One additional representation is needed to account for DM. We have 
delineated the possible representations and necessary intermediate gauge groups
needed to account for WIMP-like dark matter, proton stability, and gauge coupling unification.
Some of these models are accessible for experimental tests.

\section*{Acknowledgments}

We thank Yann Mambrini and Mikhail B. Voloshin for valuable discussions. 
The work of N.N. is supported by Research
Fellowships of the Japan Society for the Promotion of Science for Young
Scientists. The work of K.A.O. was supported in part
by DOE grant DE-SC0011842 at the University of Minnesota.

\section*{Appendix}
\appendix

\section{Dark matter candidates in SO(10)}
\label{app:DMcandidates}

Here we give a group theoretical argument to classify possible DM
candidates in SO(10) models. We basically follow the notation of
Ref.~\cite{Slansky:1981yr} in the following discussion. See also
Refs.~\cite{DeMontigny:1993gy, Mambrini:2015vna}. 

Since SO(10) is a rank-five group, we have five linearly independent
Cartan generators. We denote them by $H_i$ $(i=1,\dots, 5)$. In the dual
basis, they are expressed in terms of five-dimensional vectors as follows: 
\begin{align}
 H_1 &= \frac{1}{2}[1~2~2~1~1] ~, \nonumber \\
 H_2 &= \frac{1}{2\sqrt{3}}[1~0~0~-1~1] ~, \nonumber \\
 H_3 &= \frac{1}{2}[0~0~1~1~1] ~, \nonumber \\
 H_4 &= \frac{1}{6}[-2~0~3~-1~1] ~, \nonumber \\
 H_5 &= [2~0~2~1~-1] ~.
\end{align}
Here, $H_1$ and $H_2$ are the SU(3)$_C$ Cartan generators. $H_3$ and $H_4$
are the weak isospin and hypercharge, $T_{3L}$ and $Y$,
respectively.
$H_5$ is given by $H_5=-5(B-L)+4Y$.

Every component of an SO(10) multiplet is specified by a weight vector
$\bm{\mu}$, which is expressed by a set of five integers called Dynkin
labels as $\bm{\mu}=(\widetilde{\mu}_1, \dots, \widetilde{\mu}_5)$. Its
eigenvalues of $H_i$ are given by
\begin{equation}
 H_i(\bm{\mu}) = \sum_{j=1}^{5} \bar{h}_{ij} \widetilde{\mu}_j ~,
\end{equation}
with $H_i = [\bar{h}_{i1}, \dots , \bar{h}_{i5}]$.

The DM particle should have zero eigenvalues of $H_1$, $H_2$, and
$Q=H_3+H_4$. This condition is satisfied by the following set of weight
vectors characterized by two integers $N$ and $M$: 
\begin{equation}
\bm{\mu}_{N,M} = (-N~ N~-M~-N+M~ M)~.
\end{equation}
The hypercharge and $B-L$ charge of the weight vector are 
\begin{equation}
 Y(\bm{\mu}_{N,M}) = \frac{1}{2}(N-M)~, ~~~~~~
B-L(\bm{\mu}_{N,M}) =  N~.
\label{eq:mumnybl}
\end{equation}
We find that the $N\neq M$ cases correspond to the hypercharged DM
candidates. For $N=M$, the weight vector agrees to $\bm{\mu}_N$
discussed in Ref.~\cite{Mambrini:2015vna}.

A convenient way to determine the matter parity of the DM
candidates $\bm{\mu}_{N,M}$ is to consider congruence classes of
SO(10) \cite{Slansky:1981yr}. Two representations $R$ and $R^\prime$ of
SO(10) belong to the same congruence class if the difference between any
weights of $R$ and $R^\prime$ is always written as a linear combination
of simple roots of SO(10) with integer coefficients
\cite{Lemire:1979qd}. SO(10) has four congruence classes. Each
congruence class is characterized by a congruence number. The congruence
number of a weight vector $\bm{\mu}=(\widetilde{\mu}_1, \dots,
\widetilde{\mu}_5)$ is defined by
\begin{equation}
 c(\bm{\mu}) \equiv 2 \widetilde{\mu}_1 + 2 \widetilde{\mu}_3 -
\widetilde{\mu}_4 +\widetilde{\mu}_5 ~~~~ (\text{mod}.~4)~.
\end{equation}
For $\bm{\mu}_{N,M}$, we thus have
\begin{equation}
 c(\bm{\mu}_{N,M}) \equiv 3N+2M  ~~~~ (\text{mod}.~4)~.
\end{equation}
Then, we find that the congruence number of $\bm{\mu}_{N,M}$ is $\pm 1$
if and only if $N$ is an odd integer. Considering
Eq.~\eqref{eq:mumnybl}, we conclude that a DM particle has the odd
matter parity if it has a congruence number of $\pm 1$, while it has the
even matter parity if its congruence number is $0$ or $2$.

\section{Proton decay calculation}
\label{sec:protondecaycalc}

In this section, we describe how we calculate proton decay lifetimes in
the intermediate-scale scenarios. In these scenarios, proton decay is
induced by the exchange of the GUT-scale gauge bosons
\cite{Machacek:1979tx}. The relevant part of the SO(10) gauge
interaction is given by 
\begin{align}
 {\cal L}_{\text{int}} =\frac{g_{\text{GUT}}}{\sqrt{2}}
\bigl[
(\overline{Q})_{ar}\Slash{X}^{air}P_R (L^{\cal C})_i
+(\overline{Q})_{ai}\Slash{X}^{air}P_L (L^{\cal C})_r
+\epsilon_{ij}\epsilon_{rs}\epsilon_{abc}
(\overline{Q^{\cal C}})^{ar}\Slash{X}^{bis}P_L Q^{cj}
+\text{h.c.}\bigr] ~,
\end{align}
where
\begin{equation}
 Q=
\begin{pmatrix}
    u \\ d
\end{pmatrix}
~, ~~~~~~
 L=
\begin{pmatrix}
    \nu \\ e^-
\end{pmatrix}
~,
\end{equation}
$X$ represents the GUT gauge bosons which induce proton decay,
$g_{\text{GUT}}$ is the unified gauge coupling constant, $a,b,c$ are
SU(3)$_C$ indices, $i,j$ are SU(2)$_L$ indices, $r,s$ are SU(2)$_R$
indices, and $P_{R/L}\equiv (1\pm \gamma_5)/2$ are the chirality
projection operators. The exchange of the $X$ fields generates
dimension-six proton decay operators. These operators are expressed in a
form that respects the intermediate gauge symmetries. Between
the GUT and intermediate scales, the renormalization factors for the
effective operators are in general different among the choices of
$G_{\text{int}}$. Below the
intermediate scale, the low-energy effective theory is described by the
$\text{SU}(3)_C\otimes \text{SU}(2)_L \otimes \text{U}(1)_Y$ gauge
theory, and thus after matching the theories above and below the
intermediate scale, the prescription for the calculation is common to
all of the cases. For this reason, we first describe the calculation
below the intermediate scale. After that, we discuss each intermediate
gauge theory showing the matching conditions at the GUT and intermediate
scales as well as the RGEs between them.

In the $\text{SU}(3)_C\otimes \text{SU}(2)_L \otimes \text{U}(1)_Y$ gauge
theory, the effective Lagrangian for proton decay is generically written
as
\begin{equation}
 {\cal L}_{\text{eff}} = \sum_{I=1}^4 C_I {\cal O}_I +{\rm h.c.}~,
\end{equation}
with the effective operators given by \cite{Weinberg:1979sa,
Wilczek:1979hc, Abbott:1980zj}
\begin{align}
 {\cal O}_1&=
\epsilon_{abc}\epsilon_{ij}(u^a_{R}d^b_{R})(Q_{L}^{ci} L_{L}^{j})~,\nonumber \\
 {\cal O}_{2}&=
\epsilon_{abc}\epsilon_{ij}
(Q^{ai}_{L} Q^{bj}_{L})(u_{R}^ce_{R}^{})~,\nonumber \\
{\cal O}_{3}&=
\epsilon_{abc}\epsilon_{ij}\epsilon_{kl}
(Q^{ai}_{L} Q^{bk}_{L})(Q_{L}^{cl}
 L_{L}^{j})~,\nonumber \\
 {\cal O}_{4}&=
\epsilon_{abc}(u^a_{R}d^b_{R})(u_{R}^c e_{R}^{})~,
\label{eq:fourfermidef}
\end{align}
up to dimension six.
We then run down the coefficients to the electroweak scale. We will see
below that the coefficients $C_3$ and $C_4$ vanish in all of the cases
we consider in this paper, and thus we focus on $C_1$ and $C_2$. Their
renormalization factors are
\cite{Abbott:1980zj}
\begin{align}
  C_1(\mu) &=
\biggl[\frac{\alpha_3(\mu)}{\alpha_3
 (M_{\text{int}})}\biggr]^{-\frac{2}{b_3}}
\biggl[\frac{\alpha_2(\mu)}{\alpha_2
 (M_{\text{int}})}\biggr]^{-\frac{9}{4b_2}}
\biggl[\frac{\alpha_1(\mu)}{\alpha_1
 (M_{\text{int}})}\biggr]^{-\frac{11}{20b_1}}
C_1(M_{\text{int}})
~, \\
  C_2(\mu) &=
\biggl[\frac{\alpha_3(\mu)}{\alpha_3
 (M_{\text{int}})}\biggr]^{-\frac{2}{b_3}}
\biggl[\frac{\alpha_2(\mu)}{\alpha_2
 (M_{\text{int}})}\biggr]^{-\frac{9}{4b_2}}
\biggl[\frac{\alpha_1(\mu)}{\alpha_1
 (M_{\text{int}})}\biggr]^{-\frac{23}{20b_1}}
C_2(M_{\text{int}})
~,
\end{align}
where $b_a$ denote the one-loop beta-function coefficients for the gauge
couplings $g_a$ and $\mu$ is an arbitrary scale. We need to change the
beta-function coefficients appropriately when we across the DM mass
threshold.  Below the electroweak
scale, the QCD corrections are the dominant contribution. By using the
two-loop RGE given in Ref.~\cite{Nihei:1994tx}, we compute the Wilson
coefficients at the hadronic scale $\mu_{\text{had}}$ as
\begin{equation}
C_i(\mu_{\text{had}}) =
 \biggl[
\frac{\alpha_s(\mu_{\text{had}})}{\alpha_s(m_b)}
\biggr]^{\frac{6}{25}}\biggl[
\frac{\alpha_s(m_b)}{\alpha_s(m_Z)}
\biggr]^{\frac{6}{23}}
\biggl[
\frac{\alpha_s(\mu_{\text{had}})+\frac{50\pi}{77}}
{\alpha_s(m_b)+\frac{50\pi}{77}}
\biggr]^{-\frac{173}{825}}
\biggl[
\frac{\alpha_s(m_b)+\frac{23\pi}{29}}
{\alpha_s(m_Z)+\frac{23\pi}{29}}
\biggr]^{-\frac{430}{2001}}C_i(m_Z)~,
\end{equation}
with $i=1,2$.

In non-SUSY GUTs, the dominant decay mode of proton is
$p\to \pi^0 e^+$. The partial decay width of the mode is computed as
\begin{equation}
 \Gamma (p \to \pi^0 e^+)
=\frac{m_p}{32\pi}\biggl(1-\frac{m_\pi^2}{m_p^2}\biggr)^2
\bigl[|{\cal A}_L|^2+|{\cal A}_R|^2\bigr] ~,
\end{equation}
where $m_p$ and $m_\pi$ are the masses of the proton and the neutral pion,
respectively, and
\begin{align}
 {\cal A}_L&= C_1(\mu_{\text{had}})\langle \pi^0|(ud)_R u_L|p\rangle ~,
 \nonumber \\
 {\cal A}_R&= 2C_2(\mu_{\text{had}})\langle \pi^0|(ud)_L u_R|p\rangle ~.
\end{align}
The hadron matrix elements are evaluated with the lattice QCD
simulations in Ref.~\cite{Aoki:2013yxa}. We have
\begin{align}
 \langle \pi^0|(ud)_R u_L|p\rangle &=
 \langle \pi^0|(ud)_L u_R|p\rangle = -0.103(23)(34)~\text{GeV}^2 ~,
\end{align}
with $\mu_{\text{had}}=2$~GeV. Here, the first and second parentheses
indicate statistical and systematic errors, respectively.

\subsection{$G_{\text{int}} = \text{SU}(4)_C\otimes
\text{SU}(2)_L \otimes \text{SU}(2)_R(\otimes D)$}

For $G_{\text{int}} = \text{SU}(4)_C\otimes
\text{SU}(2)_L \otimes \text{SU}(2)_R(\otimes D)$, the dimension-six
effective operator is given by\footnote{
Note that
\begin{equation}
 \epsilon_{ij}\epsilon_{kl}\epsilon_{\alpha\beta\gamma\delta}
(\overline{\Psi^{\cal C}})^{\alpha i} P_L \Psi^{\beta j}
(\overline{\Psi^{\cal C}})^{\gamma k} P_L \Psi^{\delta l}
=
\epsilon_{rs}\epsilon_{tu}\epsilon_{\alpha\beta\gamma\delta}
(\overline{\Psi^{\cal C}})^{\alpha r} P_R \Psi^{\beta s}
(\overline{\Psi^{\cal C}})^{\gamma t} P_R \Psi^{\delta u}
= 0 ~,
\end{equation}
and thus the operator in Eq.~\eqref{eq:effop422} is the unique choice.
}
\begin{equation}
 {\cal L}_{\text{eff}} = C_{422} \cdot
\epsilon_{ij}\epsilon_{rs}\epsilon_{\alpha\beta\gamma\delta}
(\overline{\Psi^{\cal C}})^{\alpha i} P_L \Psi^{\beta j}
(\overline{\Psi^{\cal C}})^{\gamma r} P_R \Psi^{\delta s}+{\rm h.c.}~,
\label{eq:effop422}
\end{equation}
where $\alpha, \beta,\dots$ denote the SU(4) indices, and the Dirac field
$\Psi =(\Psi_L, \Psi_R)$ is defined by
\begin{equation}
 \Psi_L =
\begin{pmatrix}
 u^1_L & u^2_L& u^3_L & \nu_L \\
 d^1_L & d^2_L& d^3_L & e_L
\end{pmatrix}
~, ~~~~~~
\Psi_R^{\cal C} =
\begin{pmatrix}
 d^{\cal C}_{R1}&  d^{\cal C}_{R2}&  d^{\cal C}_{R3}& e^{\cal C}_R \\
 - u^{\cal C}_{R1} &- u^{\cal C}_{R2} &- u^{\cal C}_{R3} & -\nu _R^{\cal C}
\end{pmatrix}
~.
\label{eq:4spinordef}
\end{equation}
Here, the indices represent the SU(3)$_C$ color and ${\cal C}$ indicates
charge conjugation.
At tree level, the coefficient of the effective operator is evaluated
as\footnote{We have found that the sign of this equation is opposite to
that given in Ref.~\cite{Mambrini:2015vna}. }
\begin{equation}
 C_{422}(M_{\text{GUT}}) = - \frac{g_{\text{GUT}}^2}{2M_X^2} ~,
\end{equation}
with $M_X$ the mass of the heavy gauge field $X$. In this paper, we
neglect fermion flavor mixings \cite{FileviezPerez:2004hn} for
simplicity.

The Wilson coefficient is evolved down to the intermediate scale
using the RGE. The renormalization factor is computed to be
\cite{Munoz:1986kq}
\begin{equation}
 C_{422}(M_{\text{int}}) =
\biggl[\frac{\alpha_4(M_{\text{int}})}{\alpha_{\text{GUT}}}\biggr]
^{-\frac{15}{4b_4}}\biggl[\frac{\alpha_{2L}
(M_{\text{int}})}{\alpha_{\text{GUT}}}\biggr]^{-\frac{9}{4b_{2L}}}
\biggl[\frac{\alpha_{2R}(M_{\text{int}})}{\alpha_\text{GUT}}\biggr]
^{-\frac{9}{4b_{2R}}} C_{422}(M_{\text{GUT}})~.
\end{equation}
Then, the effective operator is matched onto the operators in
Eq.~\eqref{eq:fourfermidef}. The Wilson coefficients $C_I$ are given
by\footnote{We have fixed an error in the matching conditions given in
Ref.~\cite{Mambrini:2015vna}. }
\begin{align}
 C_1(M_{\text{int}}) &= 4 C_{422}(M_{\text{int}}) ~, \nonumber \\
 C_2(M_{\text{int}}) &= 2 C_{422}(M_{\text{int}}) ~, \nonumber \\
 C_3(M_{\text{int}}) &=  C_4(M_{\text{int}}) = 0~.
\end{align}

\subsection{$G_{\text{int}} = \text{SU}(4)_C\otimes
\text{SU}(2)_L \otimes \text{U}(1)_R$}

In the case of $G_{\text{int}} = \text{SU}(4)_C\otimes \text{SU}(2)_L
\otimes \text{U}(1)_R$, the effective Lagrangian is written as
\begin{equation}
 {\cal L}_{\text{eff}} = C_{421} \cdot 2
\epsilon_{ij}\epsilon_{\alpha\beta\gamma\delta}
(\overline{\Psi^{\cal C}})^{\alpha i} P_L \Psi^{\beta j}
(\overline{{\cal U}^{\cal C}})^{\gamma } P_R {\cal D}^{\delta } +{\rm h.c.}~,
\label{eq:effop421}
\end{equation}
with
\begin{align}
 {\cal U} &\equiv (u^1,~u^2,~u^3,~\nu) ~, ~~~~~~
 {\cal D} \equiv (d^1,~d^2,~d^3,~e) ~.
\end{align}
The GUT-scale matching condition for the operator is
\begin{equation}
  C_{421}(M_{\text{GUT}}) = - \frac{g_{\text{GUT}}^2}{2M_X^2} ~,
\end{equation}
and the renormalization factor is given by \cite{Munoz:1986kq}
\begin{equation}
 C_{421}(M_{\text{int}}) =
\biggl[\frac{\alpha_4(M_{\text{int}})}{\alpha_{\text{GUT}}}\biggr]
^{-\frac{15}{4b_4}}\biggl[\frac{\alpha_{2L}
(M_{\text{int}})}{\alpha_{\text{GUT}}}\biggr]^{-\frac{9}{4b_{2L}}}
\biggl[\frac{\alpha_{R}(M_{\text{int}})}{\alpha_\text{GUT}}\biggr]
^{-\frac{3}{4b_{R}}} C_{421}(M_{\text{GUT}})~.
\end{equation}
For the intermediate-scale matching conditions, we have
\begin{align}
 C_1(M_{\text{int}}) &= 4 C_{421}(M_{\text{int}}) ~, \nonumber \\
 C_2(M_{\text{int}}) &= 2 C_{421}(M_{\text{int}}) ~, \nonumber \\
 C_3(M_{\text{int}}) &=  C_4(M_{\text{int}}) = 0~.
\end{align}

\subsection{$G_{\text{int}} = \text{SU}(3)_C\otimes
\text{SU}(2)_L \otimes \text{SU}(2)_R \otimes \text{U}(1)_{B-L}(\otimes
  D)$}

When $G_{\text{int}} = \text{SU}(3)_C\otimes
\text{SU}(2)_L \otimes \text{SU}(2)_R \otimes \text{U}(1)_{B-L}(\otimes
  D)$, there are four independent effective operators \cite{Caswell:1982fx},
\begin{align}
 {\cal Q}_1 &= 2 \epsilon_{ij}\epsilon_{rs}\epsilon_{abc}
(\overline{Q^{\cal C}})^{a i} P_L Q^{b j}
(\overline{Q^{\cal C}})^{c r} P_R L^{ s}~, \nonumber \\
 {\cal Q}_2 &= 2 \epsilon_{ij}\epsilon_{rs}\epsilon_{abc}
(\overline{Q^{\cal C}})^{a i} P_L L^{j}
(\overline{Q^{\cal C}})^{b r} P_R Q^{c s}~,\nonumber \\
 {\cal Q}_3 &= 2 \epsilon_{il}\epsilon_{jk}\epsilon_{abc}
(\overline{Q^{\cal C}})^{a i} P_L Q^{b j}
(\overline{Q^{\cal C}})^{c k} P_L L^{ l}~, \nonumber \\
 {\cal Q}_4 &= 2 \epsilon_{ps}\epsilon_{qr}\epsilon_{abc}
(\overline{Q^{\cal C}})^{a p} P_R Q^{b q}
(\overline{Q^{\cal C}})^{c r} P_R L^{ s}~,
\end{align}
and thus the effective Lagrangian is expressed as
\begin{equation}
 {\cal L}_{\text{eff}} =\sum_{I=1}^{4} C^{(I)}_{3221} {\cal Q}_I
+{\rm h.c.}
\end{equation}
For the GUT-scale matching condition, we have
\begin{align}
 C_{3221}^{(1)}(M_{\text{GUT}}) &=
 C_{3221}^{(2)}(M_{\text{GUT}}) =
- \frac{g_{\text{GUT}}^2}{2M_X^2} ~, \nonumber \\
 C_{3221}^{(3)}(M_{\text{GUT}}) &=
 C_{3221}^{(4)}(M_{\text{GUT}}) = 0~.
\end{align}
The renormalization factors for the coefficients $C_{3221}^{(1)}$ and
$C_{3221}^{(2)}$ are given in Refs.~\cite{Munoz:1986kq,Caswell:1982fx}:
\begin{equation}
 \frac{C(M_{\text{int}})}{ C(M_{\text{GUT}})} =
\biggl[\frac{\alpha_3(M_{\text{int}})}{\alpha_{\text{GUT}}}\biggr]
^{-\frac{2}{b_3}}\biggl[\frac{\alpha_{2L}
(M_{\text{int}})}{\alpha_{\text{GUT}}}\biggr]^{-\frac{9}{4b_{2L}}}
\biggl[\frac{\alpha_{2R}(M_{\text{int}})}{\alpha_\text{GUT}}\biggr]
^{-\frac{9}{4b_{2R}}}
\biggl[\frac{\alpha_{B-L}(M_{\text{int}})}{\alpha_\text{GUT}}\biggr]
^{-\frac{1}{4b_{B-L}}}
~,
\end{equation}
for $C=C_{3221}^{(1)}$ and $C_{3221}^{(2)}$.
Then the Wilson coefficients at the electroweak scale are matched onto
those of the operators \eqref{eq:fourfermidef} as
\begin{align}
 C_1(M_{\text{int}}) &= 4 C_{3221}^{(2)}(M_{\text{int}}) ~, \nonumber \\
 C_2(M_{\text{int}}) &= 2 C_{3221}^{(1)}(M_{\text{int}}) ~, \nonumber \\
 C_3(M_{\text{int}}) &=  C_4(M_{\text{int}}) = 0~.
\end{align}

\section{Example of fine-tuning}
\label{sec:exfinetune}

To show the process of mass fine-tuning explicitly, in this section, we
consider the case of $R_\text{DM}={\bf 16}$ with $G_{\text{int}} =
\text{SU}(3)_C \otimes \text{SU}(2)_L\otimes \text{SU}(2)_R\otimes
\text{U}(1)_{B-L}$ as an example. We take $R_1={\bf 45}$, which contains
two independent SM singlet components that might develop VEVs; one is in
a $({\bf 1}, {\bf 1}, {\bf 3})$ while the other is in a $({\bf 15}, {\bf
1}, {\bf 1})$ under $\text{SU(4)}_C \otimes \text{SU(2)}_L \otimes
\text{SU(2)}_R$. We refer to these VEVs as $A_1$ and $A_2$,
respectively, and other notation is taken from Eq.~\eqref{eq:lintsc}. 
Since the components of a scalar ${\bf 16}$ have the same
quantum numbers as those of a generation of the SM fermions, we denote
them by the same symbol as for the corresponding SM fermions with a
tilde, just like the notation for sfermions in supersymmetric models. 

Let us first study the $R_{\text{DM}}^* R_{\text{DM}}^{}R_1^{}$
coupling. Since $R_1$ is the adjoint representation of SO(10), the
decomposition of this coupling in terms of the component fields has a
similar form to the gauge interaction for a ${\bf 16}$ spinor
representation. We have 
\begin{align}
 \kappa_1 {R}^*_{\rm DM} R_{\rm DM}^{} \langle R_1^{}\rangle
= \kappa_1\biggl[
 \left(-\sqrt{2}A_1-\sqrt{3}A_2\right) \widetilde{\nu}_R^*\widetilde{\nu}_R^{}
 +\left(\sqrt{2}A_1-\sqrt{3}A_2\right)
 \widetilde{e}^*_R\widetilde{e}_R^{}
+\sqrt{3}A_2 \widetilde{L}_L^* \widetilde{L}_L^{}
 \nonumber\\
 +\left(\sqrt{2}A_1+\frac{1}{\sqrt{3}}A_2\right)
 \widetilde{d}_R^*\widetilde{d}_R^{} 
 +\left(-\sqrt{2}A_1+\frac{1}{\sqrt{3}}A_2\right)
 \widetilde{u}_R^*\widetilde{u}_R^{}
 -\frac{1}{\sqrt{3}}A_2 \widetilde{Q}_L^* \widetilde{Q}_L^{}
 \biggr] ~,
 \label{eq:45mass}
\end{align}
where the contraction of the SU(3)$_C$ and SU(2)$_L$ indices is
implicit. When $A_1\neq 0$ and $A_2=0$, the mass spectrum preserves
the $\text{SU(4)}_C\otimes\text{SU(2)}_L\otimes\text{U(1)}_R$ symmetry,
while when $A_2\neq 0$ and $A_1=0$, then it is
$\text{SU(3)}_C\otimes\text{SU(2)}_L
\otimes\text{SU(2)}_R\otimes\text{U(1)}_{B-L}$ 
symmetric. If both of the VEVs have non-zero values, then the low-energy
theory is invariant under the $\text{SU(3)}_C\otimes\text{SU(2)}_L
\otimes\text{U(1)}_R\otimes\text{U(1)}_{B-L}$ symmetry. The coefficients of $A_2$ for left and right
doublets have different signs, which indicates the breaking of
left-right symmetry. Here, we choose $A_1=0 $ and $A_2=v_{\bf 45}$ to
obtain $G_{\text{int}} =
\text{SU}(3)_C \otimes \text{SU}(2)_L\otimes \text{SU}(2)_R\otimes
\text{U}(1)_{B-L}$. 

Next we consider the mass terms generated by $\lambda_2^{\bf 45}
({R}^*_{\rm DM}R_{\rm DM}^{})_{\bf 45} ({R}_2^* R_2^{})_{\bf 45}$. The
SM singlet in $R_2={\bf 126}$ transforms as $({\bf 10},{\bf 1},{\bf 3})$
under $\text{SU(4)}_C \otimes \text{SU(2)}_L \otimes \text{SU(2)}_R$,
which acquires a VEV $v_{\bf 126}$ to break $G_\text{int}$ into the SM
gauge group. According to the result in Ref.~\cite{Nath:2001yj}, the
resultant mass terms are\footnote{Note that since $ ({R}_2^* R_2^{})_{\bf 45}$ contains a 
${\bf 45}$, there is a contribution to the mass corresponding to  Eq.~\eqref{eq:45mass}
at the intermediate scale proportional to $\lambda_2^{\bf 45}$
with independent coefficients ${\tilde A_1}$ and ${\tilde A_2}$. The result shown is obtained 
from Eq.~\eqref{eq:45mass} by taking
${\tilde A_1}=\frac{\sqrt{2}}{5}v_{\bf 126}^2$ and ${\tilde A_2}=\frac{\sqrt{3}}{5}v_{\bf
126}^2$, up to an overall factor. }
\begin{align}
\lambda_2^{\bf 45}\left({R}^*_{\rm DM}R^{}_{\rm DM}\right)_{\bf 45}
\langle\left({R}^*_2R^{}_2\right)_{\bf 45}\rangle
&= \lambda_2^{\bf 45}v_{\bf 126}^2\biggl[
 - \widetilde{\nu}_R^*\widetilde{\nu}_R^{}
 +\frac{3}{5}\left(\widetilde{L}^*_L \widetilde{L}_L^{}
+ \widetilde{d}_R^*\widetilde{d}_R^{} \right)
 \nonumber\\
 &\qquad
 -\frac{1}{5}\left(
 \widetilde{e}_R^*\widetilde{e}_R^{}
 + \widetilde{u}_R^*\widetilde{u}_R^{}
+\widetilde{Q}_L^*\widetilde{Q}_L^{}\right)
 \biggr] ~.
\end{align}
Notice that the right-hand side of the expression can be grouped in
terms of SU(5) multiplets. This is expected since $v_{\bf 126}$ is
invariant under the SU(5) transformation. From the above equations, it
is found that we can ensure that only the DM component has a mass around TeV scale by
fine-tuning the parameters $M^2$, $\kappa_1$ and $\lambda_2^{\bf 45}$. 
For example, to obtain the model \DM{SA}{3221}, we can take
\begin{align}
M^2-\sqrt{3}\kappa_1 v_{\bf 45}&\sim \mathcal{O}(M^2_\text{int})~,
 \nonumber \\
M^2-\sqrt{3}\kappa_1 v_{\bf 45}-\lambda_2^{\bf 45}v_{\bf 126}^2&\sim
 \mathcal{O}(\text{TeV}^2)~. 
\end{align}
Then, $\widetilde{\nu}_R$ acquires a TeV-scale mass,
while the mass of $\widetilde{e}_R$ is $\mathcal{O}(M_\text{int})$. The
rest of the components lie around the GUT scale. On the other hand, if
we take
\begin{align}
M^2+\sqrt{3}\kappa_1 v_{\bf 45}&\sim \mathcal{O}(M^2_\text{int})~,
 \nonumber \\
M^2+\sqrt{3} \kappa_1 v_{\bf 45}+\frac{3}{5}\lambda_2^{\bf 45}v_{\bf
 126}^2 &\sim\mathcal{O}(\text{TeV}^2) ~,
\end{align}
then we can make only the $\widetilde{L}_L$ component have a TeV-scale
mass and the other components have GUT-scale masses. Thus we obtain the
\DM{SB}{3221} model.

To simplify our argument, in the above discussion, we have taken into
account only the contribution of the $M^2$, $\kappa_1$, and
$\lambda_2^{\bf 45}$ terms, and neglected that of the other terms in
Eq.~\eqref{eq:lintsc}. Even in the presence of the other contributions,
we can always perform a similar fine-tuning among the parameters to
realize desired mass spectrum for our DM models.




\clearpage
\begin{thebibliography}{99}


\bibitem{Ade:2015xua} 
  P.~A.~R.~Ade {\it et al.}  [Planck Collaboration],
  arXiv:1502.01589 [astro-ph.CO].

\bibitem{EHNOS}
H. Goldberg, Phys. Rev. Lett. {\bf 50} (1983) 1419;
J. Ellis, J.S. Hagelin, D.V. Nanopoulos, K.A. Olive
and M. Srednicki, Nucl. Phys. {\bf B238} (1984) 453.


\bibitem{Appelquist:2000nn} 
  T.~Appelquist, H.~C.~Cheng and B.~A.~Dobrescu,
  Phys.\ Rev.\ D {\bf 64}, 035002 (2001)
  [hep-ph/0012100];
  H.~C.~Cheng, K.~T.~Matchev and M.~Schmaltz,
  Phys.\ Rev.\ D {\bf 66}, 036005 (2002)
  [hep-ph/0204342];
  G.~Servant and T.~M.~P.~Tait,
  Nucl.\ Phys.\ B {\bf 650}, 391 (2003)
  [hep-ph/0206071];
  H.~C.~Cheng, J.~L.~Feng and K.~T.~Matchev,
  Phys.\ Rev.\ Lett.\  {\bf 89}, 211301 (2002)
  [hep-ph/0207125];
  M.~Kakizaki, S.~Matsumoto and M.~Senami,
  Phys.\ Rev.\ D {\bf 74}, 023504 (2006)
  [hep-ph/0605280];
  G.~Belanger, M.~Kakizaki and A.~Pukhov,
  JCAP {\bf 1102}, 009 (2011)
  [arXiv:1012.2577 [hep-ph]].

\bibitem{ArkaniHamed:2001nc} 
  N.~Arkani-Hamed, A.~G.~Cohen and H.~Georgi,
  Phys.\ Lett.\ B {\bf 513}, 232 (2001)
  [hep-ph/0105239];
  N.~Arkani-Hamed, A.~G.~Cohen, E.~Katz, A.~E.~Nelson, T.~Gregoire and J.~G.~Wacker,
  JHEP {\bf 0208}, 021 (2002)
  [hep-ph/0206020];
  N.~Arkani-Hamed, A.~G.~Cohen, E.~Katz and A.~E.~Nelson,
  JHEP {\bf 0207}, 034 (2002)
  [hep-ph/0206021];
  H.~C.~Cheng and I.~Low,
  JHEP {\bf 0408}, 061 (2004)
  [hep-ph/0405243];
  I.~Low,
  JHEP {\bf 0410}, 067 (2004)
  [hep-ph/0409025];
  J.~Hubisz and P.~Meade,
  Phys.\ Rev.\ D {\bf 71}, 035016 (2005)
  [hep-ph/0411264];
  A.~Birkedal, A.~Noble, M.~Perelstein and A.~Spray,
  Phys.\ Rev.\ D {\bf 74}, 035002 (2006)
  [hep-ph/0603077].


\bibitem{Kibble:1982ae} 
  T.~W.~B.~Kibble, G.~Lazarides and Q.~Shafi,
  Phys.\ Lett.\ B {\bf 113}, 237 (1982).

\bibitem{Krauss:1988zc} 
  L.~M.~Krauss and F.~Wilczek,
  Phys.\ Rev.\ Lett.\  {\bf 62}, 1221 (1989).

\bibitem{Ibanez:1991hv} 
  L.~E.~Ibanez and G.~G.~Ross,
  Phys.\ Lett.\ B {\bf 260}, 291 (1991);
%
  L.~E.~Ibanez and G.~G.~Ross,
  Nucl.\ Phys.\ B {\bf 368}, 3 (1992).

\bibitem{Martin:1992mq} 
  S.~P.~Martin,
  Phys.\ Rev.\ D {\bf 46}, 2769 (1992)
  [hep-ph/9207218].


\bibitem{Kadastik:2009dj} 
  M.~Kadastik, K.~Kannike and M.~Raidal,
  Phys.\ Rev.\ D {\bf 81}, 015002 (2010)
  [arXiv:0903.2475 [hep-ph]];
%
  M.~Kadastik, K.~Kannike and M.~Raidal,
  Phys.\ Rev.\ D {\bf 80}, 085020 (2009)
  [Erratum-ibid.\ D {\bf 81}, 029903 (2010)]
  [arXiv:0907.1894 [hep-ph]].

\bibitem{Frigerio:2009wf} 
  M.~Frigerio and T.~Hambye,
  Phys.\ Rev.\ D {\bf 81}, 075002 (2010)
  [arXiv:0912.1545 [hep-ph]];
%
  T.~Hambye,
  PoS IDM {\bf 2010}, 098 (2011)
  [arXiv:1012.4587 [hep-ph]].

\bibitem{Georgi:1974my} 
  H.~Georgi,
  AIP Conf.\ Proc.\  {\bf 23}, 575 (1975);
  H.~Fritzsch and P.~Minkowski,
  Annals Phys.\  {\bf 93}, 193 (1975).
  
  \bibitem{so10-2}
   M.~S.~Chanowitz, J.~R.~Ellis and M.~K.~Gaillard,
  Nucl.\ Phys.\ B {\bf 128}, 506 (1977);
   H.~Georgi and D.~V.~Nanopoulos,
  Nucl.\ Phys.\ B {\bf 155}, 52 (1979).
  
  \bibitem{GN2}
  H.~Georgi and D.~V.~Nanopoulos,
  Nucl.\ Phys.\ B {\bf 159}, 16 (1979);
  C.~E.~Vayonakis,
  Phys.\ Lett.\ B {\bf 82}, 224 (1979)
  [Phys.\ Lett.\  {\bf 83B}, 421 (1979)].
  
  \bibitem{masiero}
  A.~Masiero,
  Phys.\ Lett.\ B {\bf 93}, 295 (1980).
  
  \bibitem{ssw}
  Q.~Shafi, M.~Sondermann and C.~Wetterich,
  Phys.\ Lett.\ B {\bf 92}, 304 (1980).
  
  \bibitem{delAguila:1980at} 
  F.~del Aguila and L.~E.~Ibanez,
  Nucl.\ Phys.\ B {\bf 177}, 60 (1981).

\bibitem{Mohapatra:1982aq} 
  R.~N.~Mohapatra and G.~Senjanovic,
  Phys.\ Rev.\ D {\bf 27}, 1601 (1983).


\bibitem{Rajpoot:1980xy} 
  S.~Rajpoot,
  Phys.\ Rev.\ D {\bf 22}, 2244 (1980);
%
  M.~Yasue,
  Prog.\ Theor.\ Phys.\  {\bf 65}, 708 (1981)
  [Erratum-ibid.\  {\bf 65}, 1480 (1981)];
  J.~M.~Gipson and R.~E.~Marshak,
  Phys.\ Rev.\ D {\bf 31}, 1705 (1985);
  D.~Chang, R.~N.~Mohapatra, J.~Gipson, R.~E.~Marshak and M.~K.~Parida,
  Phys.\ Rev.\ D {\bf 31}, 1718 (1985);
  N.~G.~Deshpande, E.~Keith and P.~B.~Pal,
  Phys.\ Rev.\ D {\bf 46}, 2261 (1993);
  N.~G.~Deshpande, E.~Keith and P.~B.~Pal,
  Phys.\ Rev.\ D {\bf 47}, 2892 (1993)
  [hep-ph/9211232];
  S.~Bertolini, L.~Di Luzio and M.~Malinsky,
  Phys.\ Rev.\ D {\bf 81}, 035015 (2010)
  [arXiv:0912.1796 [hep-ph]].
  

  
\bibitem{Fukugita:1993fr} 
  M.~Fukugita and T.~Yanagida,
  In *Fukugita, M. (ed.), Suzuki, A. (ed.): Physics and astrophysics of
	neutrinos* 1-248. and Kyoto Univ. - YITP-K-1050
	(93/12,rec.Feb.94) 248 p. C. 

\bibitem{DiLuzio:2011my} 
  L.~Di Luzio,
  arXiv:1110.3210 [hep-ph].

  \bibitem{lsw}
  G.~Lazarides, Q.~Shafi and C.~Wetterich,
  Nucl.\ Phys.\ B {\bf 181}, 287 (1981).

\bibitem{Minkowski:1977sc} 
  P.~Minkowski,
  Phys.\ Lett.\ B {\bf 67}, 421 (1977);
  T.~Yanagida,
  Conf.\ Proc.\ C {\bf 7902131}, 95 (1979);
  M.~Gell-Mann, P.~Ramond and R.~Slansky,
  Conf.\ Proc.\ C {\bf 790927}, 315 (1979)
  [arXiv:1306.4669 [hep-th]];
  S.~L.~Glashow,
  NATO Sci.\ Ser.\ B {\bf 59}, 687 (1980);
  R.~N.~Mohapatra and G.~Senjanovic,
  Phys.\ Rev.\ Lett.\  {\bf 44}, 912 (1980);
  R.~N.~Mohapatra and G.~Senjanovic,
  Phys.\ Rev.\ D {\bf 23}, 165 (1981).
  

\bibitem{DeMontigny:1993gy} 
  M.~De Montigny and M.~Masip,
  Phys.\ Rev.\ D {\bf 49}, 3734 (1994)
  [hep-ph/9309312].

\bibitem{Mambrini:2015vna} 
  Y.~Mambrini, N.~Nagata, K.~A.~Olive, J.~Quevillon and J.~Zheng,
  Phys.\ Rev.\ D {\bf 91}, 095010 (2015)
  [arXiv:1502.06929 [hep-ph]].


\bibitem{Mambrini:2013iaa} 
  Y.~Mambrini, K.~A.~Olive, J.~Quevillon and B.~Zaldivar,
  Phys.\ Rev.\ Lett.\  {\bf 110}, 241306 (2013)
  [arXiv:1302.4438 [hep-ph]].

\bibitem{Babu:2015bna} 
  K.~S.~Babu and S.~Khan,
  arXiv:1507.06712 [hep-ph].


\bibitem{Bajc:2005zf} 
  B.~Bajc, A.~Melfo, G.~Senjanovic and F.~Vissani,
  Phys.\ Rev.\ D {\bf 73}, 055001 (2006)
  [hep-ph/0510139].

\bibitem{Fukuyama:2004xs} 
  T.~Fukuyama, A.~Ilakovac, T.~Kikuchi, S.~Meljanac and N.~Okada,
  Eur.\ Phys.\ J.\ C {\bf 42}, 191 (2005)
  [hep-ph/0401213].
  
  \bibitem{2-3}
   A.~Masiero, D.~V.~Nanopoulos, K.~Tamvakis and T.~Yanagida,
  Phys.\ Lett.\ B {\bf 115}, 380 (1982);
  B.~Grinstein,
  Nucl.\ Phys.\ B {\bf 206}, 387 (1982).

\bibitem{Dimopoulos:1981xm} 
  S.~Dimopoulos and F.~Wilczek,
  Print-81-0600 (SANTA BARBARA), NSF-ITP-82-07.

\bibitem{Inoue:1985cw} 
  K.~Inoue, A.~Kakuto and H.~Takano,
  Prog.\ Theor.\ Phys.\  {\bf 75}, 664 (1986);
  A.~A.~Anselm and A.~A.~Johansen,
  Phys.\ Lett.\ B {\bf 200}, 331 (1988).

\bibitem{Farrar:1978xj} 
  G.~R.~Farrar and P.~Fayet,
  Phys.\ Lett.\ B {\bf 76}, 575 (1978);
%
  S.~Dimopoulos and H.~Georgi,
  Nucl.\ Phys.\ B {\bf 193}, 150 (1981);
%
  S.~Weinberg,
  Phys.\ Rev.\ D {\bf 26}, 287 (1982);
%
  N.~Sakai and T.~Yanagida,
  Nucl.\ Phys.\ B {\bf 197}, 533 (1982);
%
  S.~Dimopoulos, S.~Raby and F.~Wilczek,
  Phys.\ Lett.\ B {\bf 112}, 133 (1982).


\bibitem{Kuzmin:1980yp} 
  V.~A.~Kuzmin and M.~E.~Shaposhnikov,
  Phys.\ Lett.\ B {\bf 92}, 115 (1980);
%
  T.~W.~B.~Kibble, G.~Lazarides and Q.~Shafi,
  Phys.\ Rev.\ D {\bf 26}, 435 (1982);
%
  D.~Chang, R.~N.~Mohapatra and M.~K.~Parida,
  Phys.\ Rev.\ Lett.\  {\bf 52}, 1072 (1984);
%
  D.~Chang, R.~N.~Mohapatra and M.~K.~Parida,
  Phys.\ Rev.\ D {\bf 30}, 1052 (1984);
%
  D.~Chang, R.~N.~Mohapatra, J.~Gipson, R.~E.~Marshak and M.~K.~Parida,
  Phys.\ Rev.\ D {\bf 31}, 1718 (1985).
  
\bibitem{Antoniadis:1987dx} 
  I.~Antoniadis, J.~R.~Ellis, J.~S.~Hagelin and D.~V.~Nanopoulos,
  Phys.\ Lett.\ B {\bf 194}, 231 (1987).
  
  
  \bibitem{flipped}
   I.~Antoniadis, J.~R.~Ellis, J.~S.~Hagelin and D.~V.~Nanopoulos,
  Phys.\ Lett.\ B {\bf 208}, 209 (1988)
  [Addendum-ibid.\ B {\bf 213}, 562 (1988)];
  J.~R.~Ellis, J.~L.~Lopez and D.~V.~Nanopoulos,
  Phys.\ Lett.\ B {\bf 292}, 189 (1992)
  [hep-ph/9207237];
  J.~R.~Ellis, D.~V.~Nanopoulos and K.~A.~Olive,
  Phys.\ Lett.\ B {\bf 300} (1993) 121
  [hep-ph/9211325];
 J.~R.~Ellis, J.~L.~Lopez, D.~V.~Nanopoulos and K.~A.~Olive,
  Phys.\ Lett.\ B {\bf 308}, 70 (1993)
  [hep-ph/9303307].


\bibitem{Slansky:1981yr} 
  R.~Slansky,
  Phys.\ Rept.\  {\bf 79}, 1 (1981).
  
\bibitem{Cirelli:2005uq} 
  M.~Cirelli, N.~Fornengo and A.~Strumia,
  Nucl.\ Phys.\ B {\bf 753}, 178 (2006)
  [hep-ph/0512090];
%
  M.~Cirelli, A.~Strumia and M.~Tamburini,
  Nucl.\ Phys.\ B {\bf 787}, 152 (2007)
  [arXiv:0706.4071 [hep-ph]];
%
%
  M.~Cirelli and A.~Strumia,
  New J.\ Phys.\  {\bf 11}, 105005 (2009)
  [arXiv:0903.3381 [hep-ph]].


\bibitem{Essig:2007az} 
  R.~Essig,
  Phys.\ Rev.\ D {\bf 78}, 015004 (2008)
  [arXiv:0710.1668 [hep-ph]].

\bibitem{Hambye:2009pw} 
  T.~Hambye, F.-S.~Ling, L.~Lopez Honorez and J.~Rocher,
  JHEP {\bf 0907}, 090 (2009)
  [Erratum-ibid.\  {\bf 1005}, 066 (2010)].

\bibitem{Hisano:2014kua} 
  J.~Hisano, D.~Kobayashi, N.~Mori and E.~Senaha,
  Phys.\ Lett.\ B {\bf 742}, 80 (2015)
  [arXiv:1410.3569 [hep-ph]].

\bibitem{Nagata:2014wma} 
  N.~Nagata and S.~Shirai,
  JHEP {\bf 1501}, 029 (2015)
  [arXiv:1410.4549 [hep-ph]].

\bibitem{Nagata:2014aoa} 
  N.~Nagata and S.~Shirai,
  Phys.\ Rev.\ D {\bf 91}, 055035 (2015)
  [arXiv:1411.0752 [hep-ph]].

\bibitem{Boucenna:2015haa} 
  S.~M.~Boucenna, M.~B.~Krauss and E.~Nardi,
  Phys.\ Lett.\ B {\bf 748}, 191 (2015)
  [arXiv:1503.01119 [hep-ph]].

\bibitem{Harigaya:2015yaa} 
  K.~Harigaya, K.~Ichikawa, A.~Kundu, S.~Matsumoto and S.~Shirai,
  arXiv:1504.03402 [hep-ph].

\bibitem{Heeck:2015qra} 
  J.~Heeck and S.~Patra,
  arXiv:1507.01584 [hep-ph].

\bibitem{Cirelli:2015bda} 
  M.~Cirelli, T.~Hambye, P.~Panci, F.~Sala and M.~Taoso,
  arXiv:1507.05519 [hep-ph];
  C.~Garcia-Cely, A.~Ibarra, A.~S.~Lamperstorfer and M.~H.~G.~Tytgat,
  arXiv:1507.05536 [hep-ph].

\bibitem{Chiang:2015fta} 
  C.~W.~Chiang and E.~Senaha,
  arXiv:1508.02891 [hep-ph].

\bibitem{Feldstein:2013uha} 
  B.~Feldstein, M.~Ibe and T.~T.~Yanagida,
  Phys.\ Rev.\ Lett.\  {\bf 112}, 101301 (2014)
  [arXiv:1310.7495 [hep-ph]].


\bibitem{Burgess:2000yq} 
  V.~Silveira and A.~Zee,
  Phys.\ Lett.\ B {\bf 161}, 136 (1985);
  J.~McDonald,
  Phys.\ Rev.\ D {\bf 50}, 3637 (1994)
  [hep-ph/0702143];
  C.~P.~Burgess, M.~Pospelov and T.~ter Veldhuis,
  Nucl.\ Phys.\ B {\bf 619}, 709 (2001)
  [hep-ph/0011335];
  H.~Davoudiasl, R.~Kitano, T.~Li and H.~Murayama,
  Phys.\ Lett.\ B {\bf 609}, 117 (2005)
  [hep-ph/0405097].

\bibitem{Deshpande:1977rw} 
  N.~G.~Deshpande and E.~Ma,
  Phys.\ Rev.\ D {\bf 18}, 2574 (1978);
  E.~Ma,
  Phys.\ Rev.\ D {\bf 73}, 077301 (2006)
  [hep-ph/0601225];
  R.~Barbieri, L.~J.~Hall and V.~S.~Rychkov,
  Phys.\ Rev.\ D {\bf 74}, 015007 (2006)
  [hep-ph/0603188];
  L.~Lopez Honorez, E.~Nezri, J.~F.~Oliver and M.~H.~G.~Tytgat,
  JCAP {\bf 0702}, 028 (2007)
  [hep-ph/0612275].

\bibitem{Arhrib:2013ela} 
  A.~Arhrib, Y.~L.~S.~Tsai, Q.~Yuan and T.~C.~Yuan,
  JCAP {\bf 1406}, 030 (2014)
  [arXiv:1310.0358 [hep-ph]];
%
  A.~Ilnicka, M.~Krawczyk and T.~Robens,
  arXiv:1508.01671 [hep-ph].


  
  \bibitem{ky}
  T.~W.~Kephart and T.~C.~Yuan,
  arXiv:1508.00673 [hep-ph].

\bibitem{Farina:2013mla} 
  M.~Farina, D.~Pappadopulo and A.~Strumia,
  JHEP {\bf 1308}, 022 (2013)
  [arXiv:1303.7244 [hep-ph]].

\bibitem{Hisano:2003ec} 
  J.~Hisano, S.~Matsumoto and M.~M.~Nojiri,
  Phys.\ Rev.\ Lett.\  {\bf 92}, 031303 (2004)
  [hep-ph/0307216];
  J.~Hisano, S.~Matsumoto, M.~M.~Nojiri and O.~Saito,
  Phys.\ Rev.\ D {\bf 71}, 063528 (2005)
  [hep-ph/0412403].

\bibitem{Shiozawa}
M.~Shiozawa, talk presented at TAUP 2013, September 8--13, Asilomar, CA,
	USA.


\bibitem{Babu:2013jba}
  K.~S.~Babu, E.~Kearns, U.~Al-Binni, S.~Banerjee, D.~V.~Baxter, Z.~Berezhiani, M.~Bergevin and S.~Bhattacharya {\it et al.},
  arXiv:1311.5285 [hep-ph].
 
\bibitem{Cline:2013gha} 
  J.~M.~Cline, K.~Kainulainen, P.~Scott and C.~Weniger,
  Phys.\ Rev.\ D {\bf 88}, 055025 (2013)
  [arXiv:1306.4710 [hep-ph]];
%
  M.~Duerr, P.~Fileviez Perez and J.~Smirnov,
  arXiv:1508.04418 [hep-ph].

\bibitem{Abe:2014gua} 
  T.~Abe, R.~Kitano and R.~Sato,
  Phys.\ Rev.\ D {\bf 91}, no. 9, 095004 (2015)
  [arXiv:1411.1335 [hep-ph]].
  



\bibitem{Akerib:2013tjd} 
  D.~S.~Akerib {\it et al.}  [LUX Collaboration],
  Phys.\ Rev.\ Lett.\  {\bf 112}, 091303 (2014)
  [arXiv:1310.8214 [astro-ph.CO]].

\bibitem{Belanger:2013xza} 
  G.~Belanger, B.~Dumont, U.~Ellwanger, J.~F.~Gunion and S.~Kraml,
  Phys.\ Rev.\ D {\bf 88}, 075008 (2013)
  [arXiv:1306.2941 [hep-ph]].


\bibitem{Abe:2015rja} 
  T.~Abe and R.~Sato,
  JHEP {\bf 1503}, 109 (2015)
  [arXiv:1501.04161 [hep-ph]].


\bibitem{Hisano:2006nn} 
  J.~Hisano, S.~Matsumoto, M.~Nagai, O.~Saito and M.~Senami,
  Phys.\ Lett.\ B {\bf 646}, 34 (2007)
  [hep-ph/0610249].

\bibitem{Lazarides:1980nt} 
  G.~Lazarides, Q.~Shafi and C.~Wetterich,
  Nucl.\ Phys.\ B {\bf 181}, 287 (1981);
%
  K.~S.~Babu and R.~N.~Mohapatra,
  Phys.\ Rev.\ Lett.\  {\bf 70}, 2845 (1993)
  [hep-ph/9209215];
%
  K.~Matsuda, Y.~Koide and T.~Fukuyama,
  Phys.\ Rev.\ D {\bf 64}, 053015 (2001)
  [hep-ph/0010026];
  T.~Fukuyama, K.~Ichikawa and Y.~Mimura,
  arXiv:1508.07078 [hep-ph].

\bibitem{Khachatryan:2014fba} 
  V.~Khachatryan {\it et al.} [CMS Collaboration],
  JHEP {\bf 1504}, 025 (2015)
  [arXiv:1412.6302 [hep-ex]];
  G.~Aad {\it et al.} [ATLAS Collaboration],
  Phys.\ Rev.\ D {\bf 90}, no. 5, 052005 (2014)
  [arXiv:1405.4123 [hep-ex]].

\bibitem{Aad:2015caa} 
  G.~Aad {\it et al.} [ATLAS Collaboration],
  arXiv:1508.04735 [hep-ex];
  CMS Collaboration [CMS Collaboration],
  CMS-PAS-EXO-12-041;
  [CMS Collaboration],
  CMS-PAS-EXO-12-042.

\bibitem{Ibe:2012sx} 
  M.~Ibe, S.~Matsumoto and R.~Sato,
  Phys.\ Lett.\ B {\bf 721}, 252 (2013)
  [arXiv:1212.5989 [hep-ph]].

\bibitem{Aad:2013yna} 
  G.~Aad {\it et al.}  [ATLAS Collaboration],
  Phys.\ Rev.\ D {\bf 88}, 112006 (2013)
  [arXiv:1310.3675 [hep-ex]].

\bibitem{Shingo:2013aja} 
  K.~Shingo,
  CERN-THESIS-2014-163;
  M.~Low and L.~T.~Wang,
  JHEP {\bf 1408}, 161 (2014)
  [arXiv:1404.0682 [hep-ph]];
  M.~Cirelli, F.~Sala and M.~Taoso,
  JHEP {\bf 1410}, 033 (2014)
  [Erratum-ibid.\  {\bf 1501}, 041 (2015)]
  [arXiv:1407.7058 [hep-ph]].

\bibitem{ams02}
A.~Kounine, talk presented at ``AMS DAYS AT CERN---The Future of Cosmic Ray
	Physics and Latest Results'', April 15--17, 2015, CERN.

\bibitem{Ibe:2015tma} 
  M.~Ibe, S.~Matsumoto, S.~Shirai and T.~T.~Yanagida,
  Phys.\ Rev.\ D {\bf 91}, no. 11, 111701 (2015)
  [arXiv:1504.05554 [hep-ph]];
  K.~Hamaguchi, T.~Moroi and K.~Nakayama,
  Phys.\ Lett.\ B {\bf 747}, 523 (2015)
  [arXiv:1504.05937 [hep-ph]].

\bibitem{Cohen:2013ama} 
  T.~Cohen, M.~Lisanti, A.~Pierce and T.~R.~Slatyer,
 {JCAP {\bf 1310}, 061 (2013)}
{[arXiv:1307.4082]};
%
  J.~Fan and M.~Reece,
  {JHEP {\bf 1310}, 124 (2013)}
  {[arXiv:1307.4400]};
%
  A.~Hryczuk, I.~Cholis, R.~Iengo, M.~Tavakoli and P.~Ullio,
  JCAP {\bf 1407}, 031 (2014)
  [arXiv:1401.6212 [astro-ph.HE]].

  
\bibitem{Abramowski:2013ax} 
  A.~Abramowski {\it et al.}  [H.E.S.S. Collaboration],
  {Phys.\ Rev.\ Lett.\  {\bf 110}, 041301 (2013)}
  { [arXiv:1301.1173]}.

\bibitem{Ackermann:2013yva} 
  M.~Ackermann {\it et al.}  [Fermi-LAT Collaboration],
  Phys.\ Rev.\ D {\bf 89}, 042001 (2014)
  [arXiv:1310.0828 [astro-ph.HE]].

\bibitem{Bhattacherjee:2014dya} 
  B.~Bhattacherjee, M.~Ibe, K.~Ichikawa, S.~Matsumoto and K.~Nishiyama,
  JHEP {\bf 1407}, 080 (2014)
  [arXiv:1405.4914 [hep-ph]].
  
\bibitem{Hisano:2004pv} 
  J.~Hisano, S.~Matsumoto, M.~M.~Nojiri and O.~Saito,
  Phys.\ Rev.\ D {\bf 71}, 015007 (2005)
  [hep-ph/0407168];
  J.~Hisano, K.~Ishiwata and N.~Nagata,
  Phys.\ Lett.\ B {\bf 690}, 311 (2010)
  [arXiv:1004.4090 [hep-ph]];
  J.~Hisano, K.~Ishiwata and N.~Nagata,
  Phys.\ Rev.\ D {\bf 82}, 115007 (2010)
  [arXiv:1007.2601 [hep-ph]];
  J.~Hisano, K.~Ishiwata, N.~Nagata and T.~Takesako,
  JHEP {\bf 1107}, 005 (2011)
  [arXiv:1104.0228 [hep-ph]].

\bibitem{Hisano:2015rsa} 
  J.~Hisano, K.~Ishiwata and N.~Nagata,
  JHEP {\bf 1506}, 097 (2015)
  [arXiv:1504.00915 [hep-ph]].

\bibitem{Billard:2013qya} 
  J.~Billard, L.~Strigari and E.~Figueroa-Feliciano,
  Phys.\ Rev.\ D {\bf 89}, no. 2, 023524 (2014)
  [arXiv:1307.5458 [hep-ph]].


\bibitem{Hill:2011be} 
  R.~J.~Hill and M.~P.~Solon,
  Phys.\ Lett.\ B {\bf 707}, 539 (2012)
  [arXiv:1111.0016 [hep-ph]];
  R.~J.~Hill and M.~P.~Solon,
  Phys.\ Rev.\ Lett.\  {\bf 112}, 211602 (2014)
  [arXiv:1309.4092 [hep-ph]];
  R.~J.~Hill and M.~P.~Solon,
  Phys.\ Rev.\ D {\bf 91}, 043504 (2015)
  [arXiv:1401.3339 [hep-ph]];
  R.~J.~Hill and M.~P.~Solon,
  Phys.\ Rev.\ D {\bf 91}, 043505 (2015)
  [arXiv:1409.8290 [hep-ph]].
  

\bibitem{Baer:2014yta} 
  H.~Baer, V.~Barger, D.~Mickelson, A.~Mustafayev and X.~Tata,
  JHEP {\bf 1406}, 172 (2014)
  [arXiv:1404.7510 [hep-ph]].
  
    \bibitem{fyl}
   M.~Fukugita and T.~Yanagida,
  Phys.\ Lett.\ B {\bf 174}, 45 (1986).


\bibitem{Lemire:1979qd} 
  F.~Lemire and J.~Patera,
  J.\ Math.\ Phys.\  {\bf 21}, 2026 (1980).


\bibitem{Machacek:1979tx} 
  M.~Machacek,
  Nucl.\ Phys.\ B {\bf 159}, 37 (1979).


\bibitem{Weinberg:1979sa}
  S.~Weinberg,
  Phys.\ Rev.\ Lett.\  {\bf 43}, 1566 (1979).

\bibitem{Wilczek:1979hc}
  F.~Wilczek and A.~Zee,
  Phys.\ Rev.\ Lett.\  {\bf 43}, 1571 (1979).

\bibitem{Abbott:1980zj}
  L.~F.~Abbott and M.~B.~Wise,
  Phys.\ Rev.\ D {\bf 22}, 2208 (1980).

\bibitem{Nihei:1994tx}
  T.~Nihei and J.~Arafune,
  Prog.\ Theor.\ Phys.\  {\bf 93}, 665 (1995)
  [hep-ph/9412325].


\bibitem{Aoki:2013yxa}
  Y.~Aoki, E.~Shintani and A.~Soni,
  Phys.\ Rev.\ D {\bf 89}, 014505 (2014)
  [arXiv:1304.7424 [hep-lat]].

\bibitem{FileviezPerez:2004hn}
  P.~Fileviez Perez,
  Phys.\ Lett.\ B {\bf 595}, 476 (2004)
  [hep-ph/0403286].

\bibitem{Munoz:1986kq}
  C.~Munoz,
  Phys.\ Lett.\ B {\bf 177}, 55 (1986).

\bibitem{Caswell:1982fx}
  W.~E.~Caswell, J.~Milutinovic and G.~Senjanovic,
  Phys.\ Rev.\ D {\bf 26}, 161 (1982).

 
\bibitem{Nath:2001yj} 
  P.~Nath and R.~M.~Syed,
  Nucl.\ Phys.\ B {\bf 618}, 138 (2001)
  [hep-th/0109116];
%
  T.~Fukuyama, A.~Ilakovac, T.~Kikuchi, S.~Meljanac and N.~Okada,
  J.\ Math.\ Phys.\  {\bf 46}, 033505 (2005)
  [hep-ph/0405300].


  
\end{thebibliography}
\end{document}